%% file: paper.tex
\documentstyle[mnextra,psfig]{mn}

\seceqnum
\nobrackets

\title[Parameter Tests Within Cosmological Simulations of Galaxy Formation]
      {Parameter Tests Within Cosmological Simulations of Galaxy Formation}

\author[Scott T. Kay et al.]
       {Scott T. Kay,$^{1}$\thanks{email: Scott.Kay@durham.ac.uk},  
	F. R. Pearce,$^{1}$ A. Jenkins,$^{1}$
        C. S. Frenk,$^{1}$ S. D. M. White,$^{2}$ 
        \newauthor P. A. Thomas$^{3}$ and H. M. P. Couchman$^{4,5}$\\
        $^1$ Physics Department, University of Durham, Science Laboratories,
        South Road, Durham DH1 3LE\\
        $^2$ Max--Planck--Institut f\"{u}r Astrophysik, Karl--Schwarzschild--Stra\ss e,
        85740, Garching, Germany\\
        $^3$ Astronomy Centre, CPES, University of Sussex, Falmer, Brighton BN1 9QJ\\
        $^4$ Department of Astronomy, Physics and Astronomy Building, University of
        Western Ontario, London, Ontario, N6A 3K7, Canada\\
	$^5$ Department of Physics and Astronomy, McMaster University, Hamilton,
	Ontario, L8S 4M1, Canada\\}

\date{\today}

\def\sfl{s}
\def\slmax{s_{\rm max}}
\def\sl0{\epsilon_{0}}
\def\Plummer{\epsilon}

\def\dt{\Delta t}
\def\dtnorm{\kappa}

\def\Zmet{Z}
\def\H0{H_0=100 \, h \, {\rm kms^{-1}Mpc^{-1}}}
\def\hMpc{h^{-1}{\rm Mpc}}
\def\hkpc{h^{-1}{\rm kpc}}

\def\hMsol{h^{-1}M_{\odot}}
\def\Gyr{\hbox{Gyr}\,}
\def\Zsol{Z_{\odot}}
\def\AP3M{{\sc ap^3m}}
\def\Nbody{{\it N}--body }
\def\Hydra{{\sc hydra }}

\def\log{{\rm log}}
\def\NSPH{N_{\rm SPH}}
\def\Ng{N_{g}}
\def\Nh{N_{h}}
\def\Mdark{M_{\rm dark}}
\def\Mgas{M_{\rm gas}}

\def\kelvin{\hbox{K}\,}
\def\Omegab{\Omega_{\rm b}}
\def\rhomin{\rho_{\hbox{min}}}

\def\etal{{et al.\thinspace}}
\def\eg{{e.g.\thinspace}}
\def\etc{{etc.\thinspace}}
\def\ie{{i.e.\thinspace}}

\def\MNRAS{MNRAS}
\def\AJ{AJ}
\def\ApJ{ApJ}
\def\ApJS{ApJS}
\def\AA{A\&A}
\def\ARAA{ARA\&A}
\def\Nature{Nature}

\def\COBE{COBE}

\begin{document}
\maketitle

\begin{abstract}
Numerical simulations of galaxy formation require a number of parameters. 
Some of these are intrinsic to the numerical integration scheme (\eg the 
timestep), while others describe the physical model (\eg the gas metallicity). 
In this paper, we present results of a systematic exploration of the effects of 
varying a subset of these parameters on simulations of galaxy formation. We use 
\Nbody and ``Smoothed Particle Hydrodynamics'' techniques to follow the evolution 
of cold dark matter and gas in a small volume. We compare a fiducial model to
24 different simulations, in which one parameter at a time is varied, focussing on
properties such as the relative fraction of hot and cold gas, and the
abundance and masses of galaxies. We find that for reasonable choices of numerical 
values, many parameters have relatively little effect on the galaxies, with the
notable exception of the parameters that control the resolution of the
simulation and the efficiency with which gas cools.
\end{abstract}


\section{Introduction}
\label{sec:intro}

Over the past two decades numerical simulations have become the most
commonly used technique for investigating the formation of structure in
the universe. Originally, \Nbody simulations were used to calculate the
clustering evolution of collisionless dark matter (\eg \cite{KlypShand};
\cite{CM83}; \cite{DEFW}). Subsequently, countless \Nbody studies have 
addressed problems ranging from the large--scale distribution of dark 
matter to the inner structure of dark matter haloes in a programme whose 
results underpin much of modern cosmology.

The success of the \Nbody approach stimulated the extension of simulation
techniques to processes involving gas. Eulerian, Lagrangian and deformable
grid methods have been developed for this purpose. One of the simplest
cosmological problem involving gas is the evolution of the intracluster
medium in rich clusters (\eg \cite{Evrard90}; \cite{THOMCOUCH};
\cite{CO94}; \cite{NFW95}; \cite{AN96}; \cite{BN98}; \cite{ENF98}). 
Since the cooling time of the gas is longer than the Hubble time in all but
the central regions of clusters, the intracluster gas may be approximated
as non-radiative. This is an important simplification that allows reliable
and repeatable results to be obtained with a variety of techniques and
numerical resolutions, as demonstrated by a recent systematic comparison
based on a simulation of a galaxy cluster in the cold dark matter (CDM)
cosmogony (\cite{FRENK99}).

Another cosmological problem that can be addressed using gas dynamics is
the evolution of cold gas clouds at high--redshift. These are responsible
for the Lyman--$\alpha$ forest lines seen in quasar spectra and their low
temperatures and moderate density contrasts ensure that cooling processes
are relatively unimportant.  Simulations of the Lyman--$\alpha$ forest in
CDM models have been very successful and have radically changed the
prevailing view of the high redshift gaseous universe (\eg Cen \etal
\shortcite{CEN94},\shortcite{CEN98}; \cite{HKWE96}; \cite{KHW99}).  Recent
work, however, has highlighted the need to take numerical effects carefully
into account when making comparisons with observations (\cite{THEUNS98}).

Simulating the formation of galaxies requires the incorporation of
radiative cooling processes into the fluid equations, enabling the gas to
achieve high enough density contrasts so that stars can plausibly form (\eg
\cite{KHW};
\cite{CO92}; \cite{ESD}; \cite{VGF1}). In general, the inclusion of such 
processes presents a formidable numerical challenge because in hierarchical
clustering theories, such as CDM, cooling is most efficient in small
objects at high redshift where cooling times are very short. As a result,
the outcome of a simulation depends on its resolution. Fortunately, the
problem is less intransigent than it might appear. It is clear that only a
small fraction of gas can have cooled into small pre--galactic objects at
high redshift. This is normally attributed to the effects of ``feedback'',
a term frequently used to denote a generic mechanism that prevents the gas
from cooling catastrophically and turning into stars
(\cite{WR}; \cite{Cole91}; \cite{WF}). Although feedback is poorly
understood, it almost certainly involves stellar energy input (supernova
explosions, galactic winds, \etc) to the intergalactic medium, and,
perhaps, even energy input from active galactic nuclei.

At present, feedback effects need to be included in simulations using a
phenomenological model. In some cases, a prescription to turn some of the
gas into stars and to add thermal or kinetic energy to the rest is explicitly 
incorporated into the simulation (\cite{K92}; \cite{NW93}; \cite{SM95}). 
In other cases (\eg \cite{VGF1}), the effect of resolution itself is used 
to cut off cooling; only gas in objects above the resolution limit of the 
simulation can cool efficiently.
 
Although resolution is the most important numerical concern in simulations
of galaxy formation, several other numerical and physical parameters
influence the outcome of a simulation. Examples of the former are the
gravitational softening or the size of the timestep; examples of the latter
are the metallicity of the gas or the baryon fraction. Often parameter
values are set in the absence of rigorous criteria. In this paper we
undertake a systematic study of a {\it limited} subset of the parameters
required in simulations of galaxy formation. We consider the way in which
such parameters affect, for example, the amount of gas that cools, the
number and masses of the galaxies that form, etc. This is not intended as
an exhaustive investigation, but rather as a guide to the sensitivity of
simulation results to these particular parameters.

We restrict our study to one particular implementation of numerical
hydrodynamics, the ``Smoothed Particle Hydrodynamics'' (SPH) technique
(\cite{GM77}; \cite{L77}).  This Lagrangian approach is well suited to
studies of galaxy formation because large dynamic variations in gas density 
and temperature are easily accommodated. In the implementation used in this
paper, our code models processes such as adiabatic compression and expansion,
shock heating and radiative cooling. For simplicity, we have chosen to
ignore the effects of photo--ionization (studied previously, for example,
by \cite{WHK97} and \cite{NS97}) and feedback. We intend to include these 
effects in a subsequent set of tests.

The rest of this paper is organized as follows. In Section~2 we outline the
numerical methods used, particularly our implementation of gravity and
hydrodynamics with radiative cooling. In Section~3 we define a ``fiducial''
simulation and describe its evolution, focussing on the distribution of
gas in the temperature--density plane and on the properties of ``galaxies.''
In Section~4 we compare this fiducial simulation to a set of simulations in
which one parameter at a time is varied. In particular, we consider the
correspondence of galaxies in different simulations as well as
properties such as their mass distribution and the distribution of
baryons within dark matter haloes. We summarise our results in Section~5 and
conclude in Section~6.

\section{Numerical Method}
\label{sec:numeric}

The data analysed in this paper were generated using
\Hydra \footnote{This code is in the public domain and can be obtained
from the Hydra Consortium at the following URL: \\
http://phobos.astro.uwo.ca/hydra\_consort/}
(\cite{CTP}, hereafter CTP95), an adaptive
particle--particle/particle--mesh $(\AP3M)$ code incorporating SPH. 
The simulations were performed on single--processor
workstations, although the same code has been implemented in
parallel on a Cray T3D (\cite{PC}). Below we briefly outline how \Hydra works.

\subsection{Gravity}
\label{subsec:gravity}

The $\AP3M$ method for calculating the gravitational forces involves
three distinct parts: the particle--mesh (PM), particle--particle (PP)
and refinement placing algorithms. The implementation used in \Hydra is
detailed in Couchman (\shortcite{COUCH}), although we outline the key
concepts below.

The PM algorithm is an efficient method for calculating forces accurately
down to scales of the order of the Nyquist wavelength of the mesh used.
The computational speed scales $\sim O(L^3 \log L)$, where $L^3$ is the 
total number of mesh cells.  The density field is discretized by smoothing the
particles on to a regular cubic mesh using the Triangular--Shaped Cloud (TSC)
kernel (\cite{HOCKEAST}). The potential of this distribution is then
obtained by performing a Fast--Fourier Transform (FFT), multiplying
by the appropriate Green's function, and doing an inverse FFT. The
potential at each mesh point is differenced to obtain the forces,
which are then interpolated back to the particle positions using the same
TSC kernel.

The PM method is augmented on small scales by the PP method, which calculates 
pairwise forces directly via a neighbour search. This is performed 
using a coarse mesh and searching only the 27 nearest cells for neighbours.
The coarser this mesh is relative to the FFT mesh, the more accurate the force 
calculation becomes at the expense of extra PP work.

On very small scales, the gravitational force must be softened not only to
avoid effects due to two--body interactions, but also to limit the need for
very small timesteps. \Hydra uses a spline (S2) softening function
which returns the Newtonian value of the force for $r \ge \sfl$,
where $\sfl$ is the softening length. For convenience we express this
value as the more common Plummer softening, $\Plummer$ (e.g. \cite{BT}),
using the equivalence $\sfl=2.34 \Plummer$. (This produces asymptotically 
matching force laws at both large and small scales.)  At late times
we employ a fixed physical softening which is the value normally quoted as
the softening for the simulation. In comoving coordinates this will
grow backwards in time proportional to $(1+z)$.  To maintain force
accuracy we do not allow the softening to grow indefinitely,
introducing a maximum value, $\slmax$, which is typically set to equal 0.3
FFT cells. Above a certain redshift (typically around $z=1$)
the softening is constant in comoving coordinates.

Since gravity is an attractive force, the particle distribution
evolves into a highly clustered state where large overdensities are
inevitable. Hence the ratio of PP to PM work rapidly increases,
leading to a substantial increase in the computational effort for each
step. The refinement placing algorithm (\cite{COUCH}) eases this by
locating highly clustered regions and recursively placing submeshes
(or refinements) over these volumes. This technique drastically
reduces the total amount of PP work per timestep, passing much of
the effort over to the faster PM algorithm.
 
\subsection{Hydrodynamics}
\label{subsec:hydro}

Hydrodynamical forces are modelled using the SPH formalism, which because
of its Lagrangian nature, is well suited to problems with a large dynamic
range in density -- a natural occurrence in cosmological simulations.
Solving the hydrodynamic equations involves estimating various fields such
as the density and pressure gradient, from the discrete particle
distribution. SPH tackles this by smoothing the particles over a finite
volume using a spherically--symmetric smoothing kernel. To cope with large 
density fluctuations, the range of the kernel is set for each particle to be 
the radius of a sphere containing approximately $\NSPH$ neighbouring particles (\cite{Wood}). 
The neighbour search is performed at the same time as the PP calculation and 
hence introduces no additional computational overhead. This procedure introduces 
a maximum SPH search length equivalent to the size of a PP chaining cell which 
leads to a minimum resolved density. For normal simulation parameters this limit 
is approximately the mean cosmological density.

Stable integration of the fluid equations requires the introduction of
an artificial viscosity term (\cite{MONGING}). The implementation used
in this study is based on the divergence of the flow, rather than the 
more usual relative pairwise velocity of particles. 

Some excellent reviews of the SPH technique are available in the literature 
(\eg \cite{MON}; \cite{BENZ}); the implementation in \Hydra is summarised 
in Thomas \& Couchman (1992, hereafter TC92; see also CTP95). The reader
is also referred to recent work by Thacker \etal (\shortcite{THACKER}) who have
investigated different implementations of the SPH force calculation, 
specifically the force symmetrization and artificial viscosity assumptions. 
Part of that study (Section 3.5) utilised the same initial conditions we
use here and examined the effects of different SPH implementations.
Our implementation is equivalent to version 1 of Thacker \etal 
(\shortcite{THACKER}).

\subsection{Radiative Cooling}
\label{subsec:cool}

Radiative cooling requires adding a sink to the energy equation
derived from the emissivity function,
\begin{equation}
\epsilon(n,T)=n^{2} \, \Lambda(T),
\end{equation}
where $n$ is the number density of atoms/ions and electrons, and $\Lambda(T)$ is 
the 
(temperature dependent) normalised cooling function. Our version of \Hydra 
uses a cooling function made up from a series of
power law fits to the optically thin radiative cooling code of
Raymond, Cox \& Smith (\shortcite{RCS}) as detailed by TC92.
This fit contains contributions from bremsstrahlung and from line
cooling due to hydrogen, helium and heavier elements with an assumed abundance 
$\Zmet$ times the solar value.  The fit is relatively good above $10^4\kelvin$
-- below this temperature the cooling function drops precipitously and
we take it to be zero. Figure~{\ref{fig:coolfn} illustrates these
curves for metallicities $\Zmet/\Zsol=0.0,0.5,1.0$. The $\Zmet=0.5\Zsol$
cooling function of Sutherland \& Dopita (\shortcite{SUTDOP}) is overlaid. 

\begin{figure}
\centering
\centerline{\psfig{file=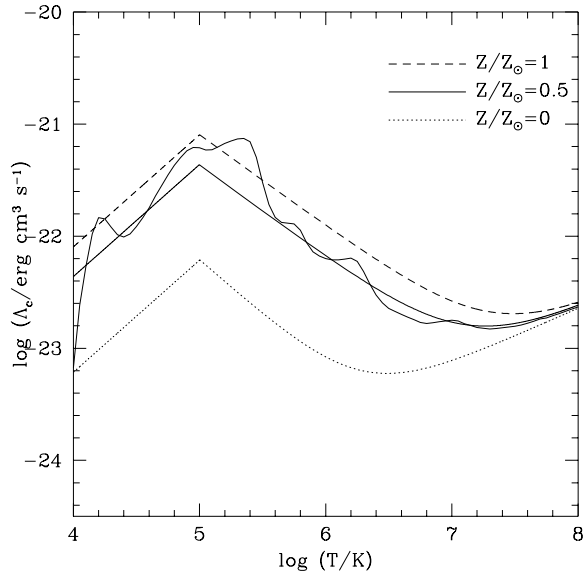,height=8cm}}
\caption{The fits used in \Hydra for the normalised cooling function,
$\Lambda(T)$. The three curves illustrate this for three values of the
metallicity: $\Zmet/\Zsol=0.0,0.5,1.0$.The function is only plotted
for temperatures down to $10^{4}\kelvin$ since this is the limit where
\Hydra switches off the cooling. (Physically, negligible thermal energy
is lost below this value.) Overlaid on the $\Zmet=0.5\Zsol$ line is the
interpolated curve from the tabulated values given by Sutherland \& Dopita (1993).}
\label{fig:coolfn}
\end{figure}

\subsection{Numerical Integration}
\label{subsec:numint}

The numerical integration algorithm is described in CTP95, who evaluated
several algorithms and settled for a simple PEC (Predict, Evaluate,
Correct) scheme. This has the advantage of keeping storage to a minimum while also
allowing arbitrary changes in the timestep, if the force changes abruptly in
time (as it can do in the vicinity of shocks).  \Hydra evaluates the
timestep, $\dt$, such that $\dt =
\dtnorm \, {\rm min}(0.4 \dt_{v}, 0.25 \dt_{a}, 0.0625 \dt_{H})$, where 
$\dt_{v}=\Plummer/v_{\rm max}$, $\dt_{a}=(\Plummer/a_{\rm max})^{1/2}$
and $\dt_{H}=1/H$, with $v_{\rm max}$ being the current largest speed,
$a_{\rm max}$ the current largest acceleration and $H$ the Hubble parameter.
The value of $\dtnorm$ is a normalisation constant typically set to unity for 
adiabatic simulations (TC92). A Courant--style condition is implicit in
$\dt_{v}$, since the minimum SPH smoothing length for the gas is set by
the softening. Note that a single timestep is
used to advance all particles. In principle, a further gain in efficiency
could be obtained by implementing individual timesteps for each particle.

Although the cooling timescale is often shorter than the dynamical (or
sound--crossing) timescale, we do not shrink the timestep to follow cooling
explicitly. Instead, we apply the cooling condition, assuming a constant density,
at the end of the timestep. Thermal energy is removed such that
\begin{equation}
\int\limits_{e_i}^{e_i-\Delta e_i} {de_i \over \Lambda(T_i)} = - {n_i^{2} \over \rho_i} \, \Delta t,
\end{equation}
with subscript $i$ denoting the $i^{th}$ particle and  $e$, its specific thermal 
energy. This results in stable and physically reasonable behaviour, even for
relatively long timesteps.

\section{The Fiducial Simulation}
\label{sec:fiducial}

\subsection{Initial Conditions}
\label{subsec:ics}

Throughout this paper we examine the effects of varying various numerical
and physical parameters relative to a fiducial simulation.
Our fiducial initial conditions were chosen to correspond to the
standard cold dark matter model (SCDM). Thus we choose
the cosmic density parameter, $\Omega=1$, the cosmological constant,
$\Lambda=0$ and the Hubble constant, $h=0.5$ \footnote {$\H0$}.
Although this model is now out of favour (for example, its power spectrum 
cannot simultaneously fit both the large--scale \COBE \ normalisation
and the amplitude of small--scale galaxy clustering), it is a well studied 
model with moderately large power on all scales relevant to galaxy formation. 
In any case, the precise form of the initial conditions is not important for 
the tests carried out here.

The initial mass
distribution was generated as described by
Efstathiou \etal (\shortcite{EDFW}), by perturbing a uniform
distribution of particles using the Zel'dovich approximation.  The
amplitudes of the Gaussian random field were set using Bond \&
Efstathiou's (1984) approximation to the CDM power spectrum.  The
normalisation was set so that $\sigma_8$, the $rms$ of linear mass 
fluctuations in top--hat spheres of comoving radius, $8 \, \hMpc$,
matches the value derived from the local abundance of rich clusters. 
We adopted $\sigma_8 = 0.6$ (\cite{WEF}; \cite{VL}; \cite{ECF}).  
The size of the simulation box (in comoving
coordinates) was fixed at $10 \, \hMpc$, with a mesh resolution of
$L=64$ (implying a cell--length of $156.25 \, \hkpc$). Although the box
is not big enough to follow the evolution of large scale structure 
accurately (waves on the scale of the box are not evolving linearly at the
final time), this is not important since we are not attempting to compare 
our results directly to observations -- the volume is far too small for 
such purposes.

Table \ref{tab:fidparam} summarises the fiducial values of all the
parameters that we study in this paper. We represent the mass
using $32^{3}$ gas and $32^{3}$ dark matter particles.  The initial
softening was set to 0.3 FFT cells (equivalent to a comoving Plummer
softening of $ \sim 20 \ \hkpc$), but after $z \sim 1$ it switched to a
fixed physical softening of $10 \ \hkpc$. The number of neighbours
over which the SPH algorithm smooths 
was set to $\NSPH=32$.  The simulation was run from an
initial epoch, $z_{i}=24$, to the present ($z=0$), with the data output
at regular intervals.

Regarding the physical parameters, we chose to set the baryon fraction
equal to $\Omegab=0.06$, consistent with the results from nucleosynthesis
calculations (\cite{CST}). The global gas metallicity was set to $\Zmet =
0.5 \Zsol$, similar to the value obtained from spectral observations of
intracluster gas (\eg \cite{MLATFMKH}). The gas was given a cold start,
with the initial temperature set to $100 \kelvin$.

\input tab_fidparam

\subsection{The Baryon Phase Diagram at ${\mathbf z=0}$}
\label{subsec:phase}
\begin{figure*}
\centering
\centerline{\psfig{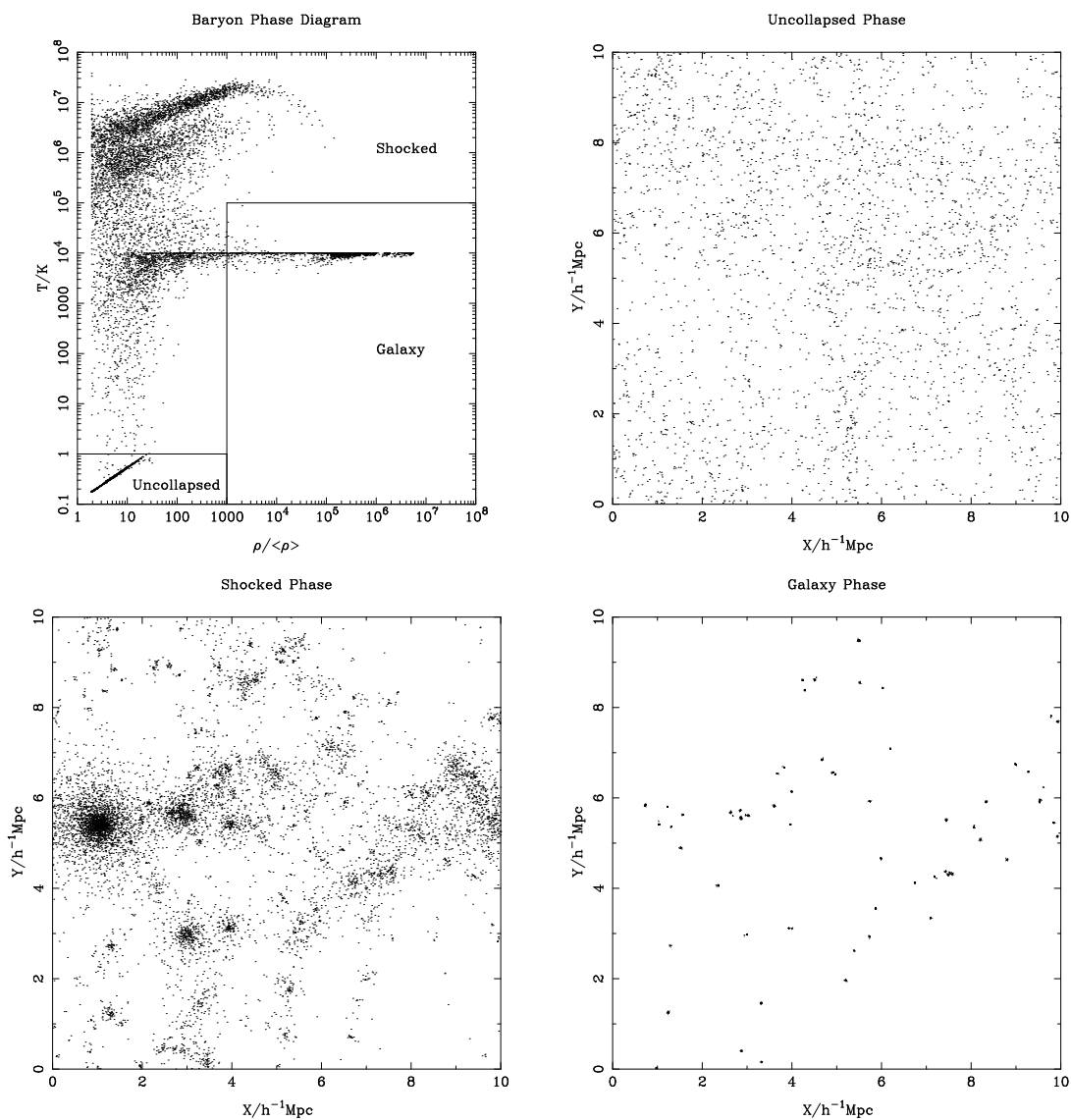}}
\caption{(top--left) The temperature--density phase diagram of all
baryons in the fiducial simulation at the final output time
($z=0$). The density is displayed in units of the mean gas density and
the temperature in Kelvin.  Positions of the gas particles obeying the
various selection criteria defined in the text are shown in the other
three plots, specifically the uncollapsed, shocked and galaxy phases. It is
evident that each phase shows a considerable difference in structure:
the uncollapsed phase particles are homogeneously distributed, the shocked
phase particles show large fluctuations in density and the galaxy
phase is exclusively made up of tight clumps of particles.}
\label{fig:phase}
\end{figure*}

The two objects that we study in this paper are the dark
matter haloes and the cold, dense clumps of gas that we define as galaxies.
All our comparisons are carried out at $z=0$. The selection criteria
used to catalogue the haloes and the galaxies are described in detail in
section~\ref{subsec:select}, but as motivation for the galaxy selection
procedure, we initially develop an understanding of the
correspondence between the spatial distribution and the thermal evolution
of the baryons. Figure~\ref{fig:phase} illustrates the
temperature--density distribution of the gas at $z=0$, spanning $7$
orders of magnitude in density and $9$ orders of magnitude in
temperature. We have split the distribution into three {\it phases},
which we label as the {\it uncollapsed} phase, the {\it shocked} phase and the
{\it galaxy} phase. Accompanying this figure are three dot--plots,
showing the corresponding spatial distribution of each phase.

\subsubsection{The uncollapsed phase}
\label{subsec:coldphase}
We define the uncollapsed phase as the regime in which gas particles
have $ T \leq 1 \kelvin$ and $\rho < 10^{3}\left<\rho\right>$. The
boundaries for this phase are chosen such that they enclose the particles that 
lie on the power law relation, which is an adiabat. These 
particles are still largely in free expansion and are, on average, 
cooling adiabatically with $T\propto(1+z)^2$. Spatially, the particles are homogeneously
distributed, as a result of occupying a small range in density.
The hard lower limit on this quantity (visible for all temperatures)
is artificial, imposed by the algorithm due to the maximum distance 
it can search for $\NSPH$ neighbours. For the fiducial simulation, 
$\rhomin\sim 1.9 \left<\rho \right>$, clearly an undesirably large value. 
As a result, the low density gas is forced to unphysically high values 
(the gas should have $\rho < \left<\rho\right>$). The 
latest version of \Hydra now properly accounts for low density gas and
we have checked that this makes no significant difference
to the properties of the high density material.

\subsubsection{The shocked phase}
The shocked gas phase contains particles with $\rho < 10^3 \left<\rho\right>$
and $T > 1 \kelvin$ as well as all particles with $T > 10^5 \kelvin$. 
Gas entered this phase due to the collapse and violent relaxation of gravitationally
unstable regions, in which the bulk kinetic energy of the gas was transferred,
via shocking, into heat. As a result, the spatial distribution of this phase
exhibits large fluctuations in density: halo atmospheres lie in this regime and the 
shocked gas is largely absent from the voids. 

\subsubsection{The galaxy phase}
The particles in the galaxy phase lie in the regime, $
\rho \ge 10^3 \left<\rho\right>$, and $T \le 10^5\kelvin$. The
strong horizontal line at $10^{4}\kelvin$ is due to the fact that the
cooling rate drops to zero at this temperature.
Particles above a temperature of $10^{5}\kelvin$ must
lose a significant fraction of their thermal energy before reaching
the main locus, a transition made possible by the large contributions from 
hydrogen and helium line cooling at these temperatures.  
The plot of their positions highlights the fact that cold, dense gas 
forms very tight clumps, with spatial extents of the order of the gravitational
softening. To select ``galaxies'', we look at the particles 
that lie in this region of the phase diagram. 

\begin{figure}
\centering
\centerline{\psfig{file=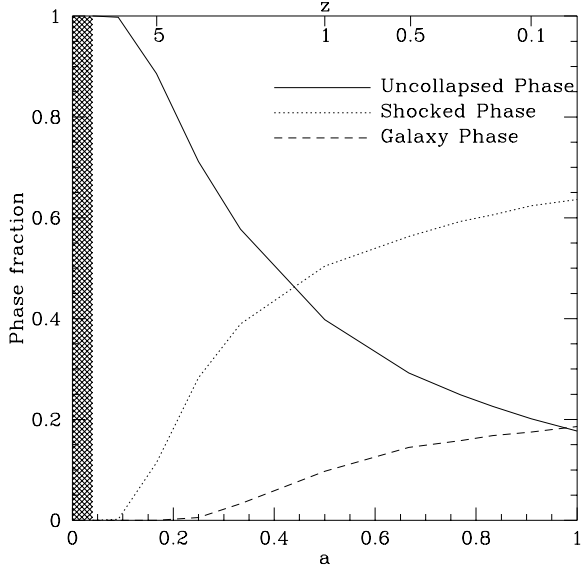,height=8cm}}
\caption{The evolution of the uncollapsed, shocked and galaxy phases, plotted as
the fraction of gas particles in each phase (with respect to the total
mass of gas particles), as a function of the
expansion factor, $a=(1+z)^{-1}$.  The shaded region marks the time
before the start of the simulation ($a=0.04$). Redshift is indicated along the
top of the figure.}
\label{fig:phase_evol}
\end{figure}

\subsection{Evolution of the Baryon Phase Diagram}
\label{subsec:evolphase}

To illustrate the global evolution that led to the baryon phase
diagram of Figure~\ref{fig:phase}, we plot the mass fraction 
in each phase in Figure~\ref{fig:phase_evol}, as a function of the expansion 
factor (normalised to the present value). All phase boundaries are fixed,
with the exception of the uncollapsed phase -- the temperature threshold
is modified to $T< (1+z)^{2}\kelvin$, accounting for the adiabatic expansion of the
uncollapsed gas. Initially, all the gas is in this phase, 
with a temperature of $100 \kelvin$.  As the simulation progresses, some of 
the gas transfers to the shocked phase, as the first virialized
haloes become resolved ($z \sim 10$). Gas does not start to enter the
galaxy phase (forming the first objects) until later ($z \sim
3$). Both the shocked and galaxy phases continue to accrete more and more
mass as the uncollapsed phase becomes progressively depleted.
By the present day, $\sim 18\%$ of the gas is still in the
uncollapsed phase 
whilst the shocked gas phase contains about $64\%$ of the gas mass and the
remaining $18\%$ has cooled into the galaxy phase.

\begin{figure}
\centering
\centerline{\psfig{file=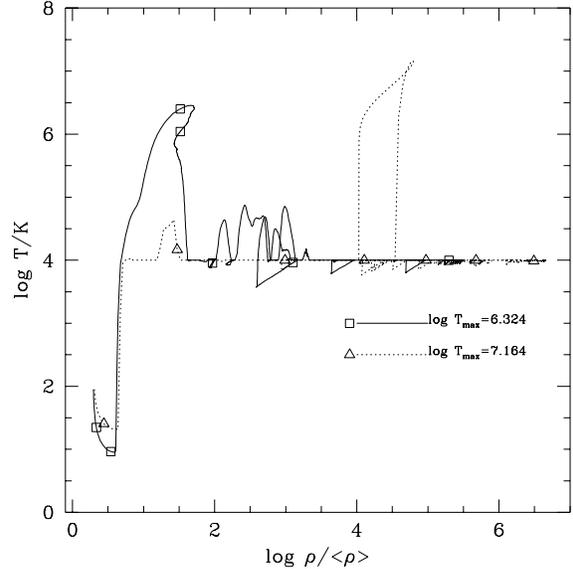,height=8cm}}
\caption{Two examples of particle trajectories in the $\rho-T$ plane,
from the initial epoch ($z=24$) to the present day. The legends
depict their maximum temperatures.  The solid path illustrates the
particle that has the highest temperature on initial heating
(\ie the first time it crosses the $10^4\kelvin$ boundary), whereas the
dotted path illustrates the evolution of a particle which is only
heated to $10^4\kelvin$. Marked points indicate redshifts 
$z=(10,5,3,2,1,0.5,0.1)$.}
\label{fig:tr_2}
\end{figure}

Another important question we have addressed is to determine the precise
trajectories in the $\rho-T$ plane, of the particles that finally
end up in the galaxy phase. Figure~\ref{fig:tr_2} illustrates the
two main paths that the galaxy particles take, for the whole duration
of the fiducial simulation. The marked points indicate the 
redshifts $z=(10,5,3,2,1,0.5,0.1)$ for each trajectory.

The first example (solid line) shows the expected evolution of a galaxy
particle -- it becomes initially shock heated to over $10^6\kelvin$ 
before cooling back down on to the $10^4\kelvin$ boundary and,
on average, progressively increasing in density until the final time. 
The second example
(dotted line) takes a different path -- again, the particle is shock heated, 
but only as far as $10^4\kelvin$. Both particles undergo a series of
further heating events and, notably, the second one reaches its largest
temperature at some later stage when a major galaxy merger takes place.

Quantitatively, we find that only $\sim 11\%$ of the particles which end
up in the galaxy phase were initially heated above $10^5\kelvin$ for a minimum
of one timestep ($\sim 32\%$ initially reached temperatures above $12500\kelvin$). 
Most of the galaxy mass reaches high density in the smallest haloes close to
the resolution limit of the simulation. Consequently, their very short cooling 
times causes them to be heated above the $10^4 \kelvin$ locus and subsequently cool 
back down within one timestep.

\subsection{Selection Criteria}

\label{subsec:select}
\begin{figure}
\centering
\centerline{\psfig{file=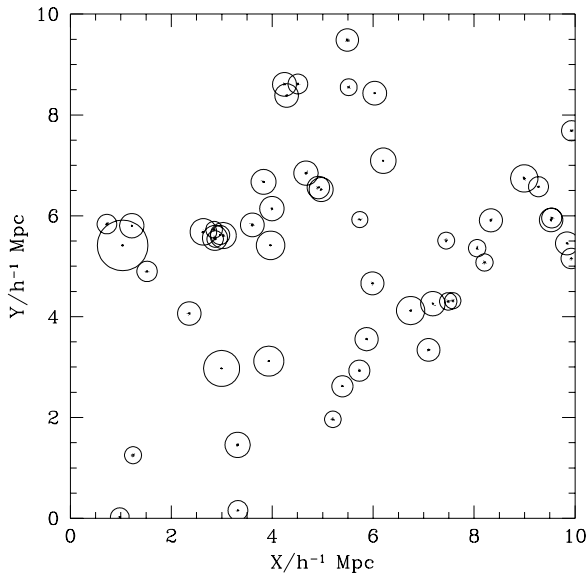,height=8cm}}
\caption{Projected positions of the fiducial set of 53 galaxies. The dots
show the particles that are linked together by the group finder and the
circles illustrate the size of the galaxies, centred on the median 
position of each galaxy with radii proportional to the cube root of the 
galaxy mass.}
\label{fig:galaxies}
\end{figure}

To select the ``galaxies'', we initially extracted the subset of the gas
particles that lie in the defined galaxy phase (Figure~\ref{fig:phase}).
This was then run through a friends--of--friends group finder
(\cite{DEFW}), which links particles together to form groups until no
unlinked particles remain that are closer to a group than the linking
length, $l=b\overline{n}^{\,-1/3}$, where $\overline{n}$ is the global mean
number density. We used the value $b=0.1$ ($l\sim 30 \hkpc$), which gives a
linking length of the order of the final S2 softening length. Since the
clumps are very tight and are well separated, the output of the group
finder is insensitive to variations in the choice of $b$: changes of the
order of a few percent in mass were noted when varying $b$ by a factor of
two in each direction.

Once this catalogue was generated it was truncated by
throwing away objects containing fewer than $\NSPH$ particles.
Objects below this mass are poorly defined because the SPH algorithm
does not sample their properties well. For the fiducial simulation, 
16 groups out of 69 were discarded in this manner, making up $\sim 4\%$ 
of the mass in the galaxy phase. This constitutes most of the residual material 
in the galaxy phase that is not part of the final galaxy catalogue.

We have also checked the effect of varying our defined limits of the
galaxy phase. Since we make use of the  total mass of gas in this phase, it is
important to ensure that we have selected a region within which most of the
particles are part of objects in the galaxy catalogue.  The most
important boundary is the lower limit to the gas density ($
\rho/\left<\rho\right> = 10^3$). Since the $10^{4}\kelvin$ feature is
visible down to the lowest densities, there are particles that have
gone through the cooling process that lie outside our galaxy phase.
Lowering the density limit by an order of magnitude caused only a 3\%
increase in the total number of particles grouped into galaxies. On the
other hand, the fraction of the galaxy phase present in the final galaxy
catalogue dropped from $96\%$ to $80\%$.
Reducing the density by the same amount again caused no further change in the
former quantity but the completeness again decreased to $56\%$, indicating
that at these low densities there is contamination from the diffuse halo gas.  
The temperature boundary for the galaxy phase includes all relevant
particles for $\rho > 10^3 \left< \rho \right>$, since the
hotter particles are still cooling
down. As a check however, we lowered the upper limit on the
temperature from $10^5\kelvin$ to $12000\kelvin$. This produced a change of
less than $1\%$ to the completeness and lowered the number of galaxies
in the catalogue from 53 to 52.

The positions of the fiducial galaxies are illustrated in Figure~\ref{fig:galaxies}. 
Here we plot the projected positions of the particles
linked together by the group finder. The circles illustrate the size
of the objects, centred on the median position of the linked particles 
in each group, with radii proportional to the cube root of the number 
of particles linked.

\begin{figure}
\centering
\centerline{\psfig{file=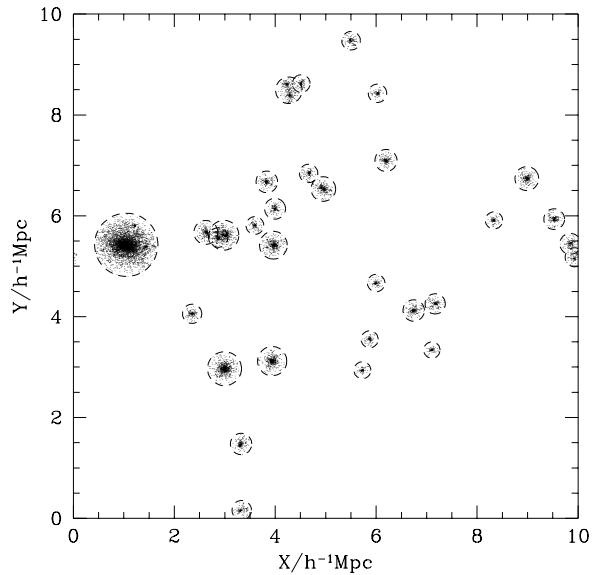,height=8cm}}
\caption{Projected positions of the dark matter particles in the fiducial
simulation selected by the halo finder, Only haloes with more than 100
particles are shown. The circles are centred on each halo's centre of
mass, and their radii are set equal to the distance from the centre to the
most outlying particle of the halo. There are 31 haloes in total in
the fiducial catalogue.}
\label{fig:haloes}
\end{figure}

The dark matter haloes were identified by using a spherical overdensity
algorithm (\cite{LC94}) on the positions of the dark matter
particles in the fiducial simulation endstate.  The
algorithm makes an estimate of the local density which is then used to
find the centre of overdense regions. Successive spheres are placed around
this centre with growing radii, until the mean internal density equals some
value -- we assume the usual value of 178 times the mean density, as
predicted by the spherical collapse model for the virial radii of
haloes in an $\Omega=1$ universe.  Iterations are performed, with the
centre redefined as the centre of mass of the object, until the
difference becomes small. Substructure within haloes was ignored for
the purposes of this paper. The halo finder produces a catalogue of all
haloes with more than $\Nh = 100$ particles in the fiducial case, 
corresponding to a mass of $\sim 8 \times 10^{11} \hMsol$.
Since the dark matter dominates the dynamical evolution, 
a lower limit in halo mass (rather than particle number) was applied 
when comparing different simulations. Hence, for simulations with poorer 
mass resolution, 
the lower limit on the number of particles was reduced in order to
keep the halo mass threshold constant. This limit never went below 10
particles, roughly the lowest number that can be trusted (\cite{EDFW88}).

The output of the halo finder for the fiducial simulation is
illustrated in figure~\ref{fig:haloes}. The circles are centred on the
halo centre of mass and have radii equal to the virial radius
of each halo.
The dots represent the positions of all the dark matter
particles within the virial radius of each halo. The largest halo has a
total mass of $\sim 5 \times 10^{13} \hMsol$, which corresponds to a
virial temperature of $\sim 10^7\kelvin$.

\begin{figure}
\centering
\centerline{\psfig{file=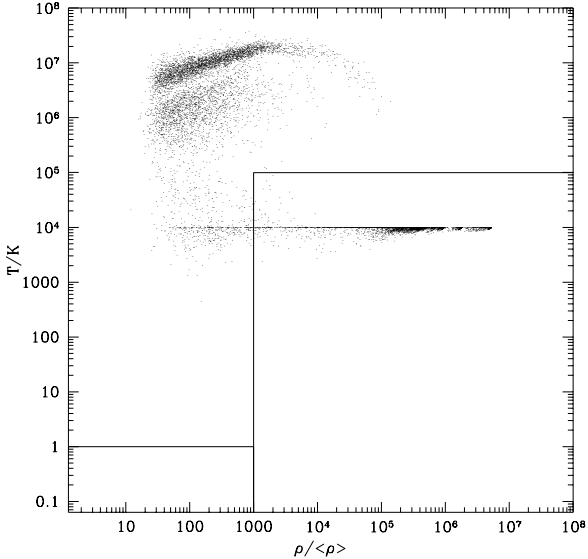,height=8cm}}
\caption{Phase diagram of all the gas particles within the selected
halo virial radii in the fiducial simulation. Overlaid are the
boundaries which we have defined in order to split the gas into 
separate phases.}
\label{fig:hfph}
\end{figure}

The $\rho-T$ distribution of the baryons located within the halo virial radii 
for the fiducial simulation are shown in Figure~\ref{fig:hfph}.  The range of 
temperatures for the majority of the hot phase gas agrees with the corresponding 
range of virial temperatures of the haloes for the mass range of the simulation
(from $\sim 5 \times 10^5-10^7 \kelvin$). Note that only a negligible amount
of cold gas falls within the halo boundaries.

\section{Simulation Comparisons}
\label{sec:simcomp}

\subsection{Global Features}
\label{subsec:global}

\input tab_select2

The complete set of simulations that we analysed are listed in Table~\ref{tab:select}, 
split into three sections. The first section
includes simulations in which features other than simple parameter
values were varied. The second section contains simulations in which numerical
parameters (i.e. those that affect the algorithm) were varied. Finally, the 
third section contains simulations in which
the physical parameters were altered. For most of the parameters,
there is a set of three simulations: the fiducial case and two
additional simulations with the appropriate parameter varied about
the fiducial value. The columns indicate the comparison number (13 in
total); the alteration made to the fiducial parameter set; the
fraction of particles identified in the uncollapsed, shocked and galaxy phases respectively;
the number of galaxies in the final catalogue; the number of haloes
in the final catalogue; and the {\it completeness}, which we define as
the fraction of particles in the galaxy phase that appear in the
final object set. For discussion, we label the sections {\it general},
{\it numerical} and {\it physical} respectively.

\subsubsection{General Comparisons}

The first comparison was designed to probe effects that might be present 
due to the fact that runs were performed on different computer 
architectures. The fiducial simulation was run on a Sun
Sparcstation, whereas most of the comparison runs
were performed on a Dec Alpha. Hence, we have
analysed an identical simulation to the fiducial case that was
executed on the latter platform. Differences arise from
the system--specific changes made to the code by the
compiler. Clearly, this change can introduce variations in the
phase fractions at the level of $\sim 1-2\%$. Furthermore, the number
of galaxies in this comparison differs by 2 and the number of haloes
differs by 1. 

Comparison 2 was performed in order to investigate the numerical
effects introduced by perturbing the particles slightly from their initial
positions. We set up two simulations to test this.  The
first also tested the periodicity of the algorithm, by
translating the particles by a fixed amount (in this case 1 FFT
cell) before starting the simulation, then moving them back at the
final time.  The second was performed in order to examine how
small inaccuracies in the initial positions affect the results, by
adding a random perturbation (tiny compared to the initial displacements)
to each particle position in the initial conditions. Variations in the
phase fractions and number of objects are similar to the first comparison,
hence the two comparisons set a baseline for the size of variations
that can be introduced solely by noise. Simulations that only differ
by this amount should be viewed as being indistinguishable from the
fiducial case.

For comparison 3, we have added a run that uses a cooling function in
tabulated form, as given by Sutherland \& Dopita (\shortcite{SUTDOP}),
rather than a series of power--law fits. The two cooling functions
used are shown in Figure~\ref{fig:coolfn} for the fiducial
metallicity, $Z/Z_{\odot}=0.5$. The power--law fits the tabulated function
relatively well for the range of relevant temperatures ($10^4-10^8\kelvin$), 
although the latter shows a significant enhancement in $\Lambda(T)$ around 
$10^5\kelvin$, (as much as a factor of 3) due to helium line cooling.
However, as Table~\ref{tab:select} shows, there only appears to be a
slight enhancement in the amount of gas in the galaxy phase.

The final comparison that was made in this section was between the
fiducial simulation and the version of \Hydra used by Pearce \etal
(\shortcite{VGF1}). The latter contains a change in the SPH algorithm
to reduce cooling in large haloes (see also \cite{THACKER}). It is
based on a {\it multiphase} approach, in which the cold, dense particles
are decoupled from the hot medium: gas above $10^{5}\kelvin$ does not see 
gas below $12000\kelvin$ for the purposes of calculating the gas density
and forces; for all other gas particles, they are calculated
in the standard way. This procedure ensures that infalling cold
diffuse particles still feel the halo's outer accretion shock and allows
hot haloes to cool at a rate determined by only the hot gas and galaxies 
to dissipate energy, merge and feel drag as they move around the halo 
environment. The effects of this alteration on the phase distribution are evident in
the table: the fraction of gas particles in the galaxy phase is
reduced by $3\%$ (with $\sim 2\%$ due to the largest object)
while the fraction in the shocked phase increases to compensate.

\subsubsection{Varying Numerical Parameters}

The first numerical parameter we varied (comparison 5) was the initial (maximum) 
softening, $\slmax$. Increasing $\slmax$ moves the changeover
from a comoving to a physical softening to higher redshift. We
doubled the value of the maximum softening to $\slmax=0.6$, so that
for the same final softening length, the changeover now occurred
at $z \sim 3$ rather than at $z \sim 1$. More importantly, the softening
now has a larger comoving value (equivalent to a Plummer softening of $
\sim 40 \ \hkpc$) before the changeover. The galaxies that form before
this epoch are more loosely bound (increasing their effective
volume), and therefore are susceptible to greater variations in their mass, due to
ram--pressure stripping, tidal disruptions or accretion as they move
through the hot halo gas. No significant changes in the final phase
fractions are present however.

The timestep normalisation value, $\dtnorm$ (comparison 6) was varied
by a factor of 2 about the fiducial value of 1. This effectively
doubles or halves the length of each timestep, causing the simulation
to take roughly twice or half as many timesteps to run to $z=0$.
Doubling the fiducial value introduces inaccuracies
in both the positions and velocities of the particles.
The phase fractions show that there is significantly more shocked gas in
the $\dtnorm=2$ simulation than in the fiducial run, at the expense
of the uncollapsed gas. 
The increased efficiency at which the gas is heated can be explained
by the fact that the larger timesteps in the $\dtnorm=2$ run increases the 
error in particle positions and velocities, which leads to larger values in the 
velocity divergence, increasing the efficiency of gas heating from viscous 
interactions.

As was hoped, the value $\dtnorm=0.5$ had no significant effect on the endstate 
properties of the galaxies and haloes, vindicating CTP95's choice of $\dtnorm=1$.

Comparison 7 consisted of halving and doubling the value of the final
softening length about the fiducial value, $\sl0 = 10 \ \hkpc$.  Changing
the value of $\sl0$ only has an effect once the physical softening length
has dropped below the maximum defined by $\slmax$. For $\sl0 = 5 \ \hkpc$
this occurs at $z=3$ whilst for $\sl0 = 20 \ \hkpc$ the softening is always
comoving. Doubling the fiducial value of $\sl0$ has no significant effect
on the final number of galaxies and the fraction of gas in each
phase. However halving the final softening length systematically decreases
the number of galaxies by $\sim 15\%$, with the residual material still in
the shocked phase. This might be due to the effects of two--body relaxation
artificially heating gas particles (\cite{SW}). For a dark matter halo with
50 particles that forms at $z=0$, the two--body relaxation times are
approximately $10.4, 7.5 \ \& \ 5.9 \ \Gyr$ for $\sl0=20,10 \ \& 5 \ \
\hkpc$ respectively. For $\sl0=5 \ \hkpc$, the relaxation time is less than
half the age of the universe.

The value of $\NSPH$ was varied in comparison 8. Since this parameter sets
the resolution at which the density field is evaluated, large overdensities
are progressively smoothed out as the value of $\NSPH$ increases. It then
becomes increasingly difficult to form galaxies, since the emissivity is
proportional to the square of the gas density. On the other hand, using a
smaller value of $\NSPH$ increases the noise arising from the discrete
nature of the mass distribution. Thus a compromise in the value is sought.
It is common to assume $\NSPH=32$, which is our fiducial value, and this
comparison looks at the effects introduced by changing this number by a
factor of two. The number of galaxies in each simulation clearly shows the
expected trend, with more galaxy phase gas and more galaxies if $\NSPH=16$
than if $\NSPH=64$. This material comes from the shocked phase as the 
uncollapsed gas fraction is more or less constant in all cases. The significant drop
in completeness for the $\NSPH=64$ simulation is due to 11 groups being
discarded because they are below the imposed resolution threshold (\ie for
these galaxies, $N_g<\NSPH$).

\begin{figure*}
\centering
\centerline{\psfig{file=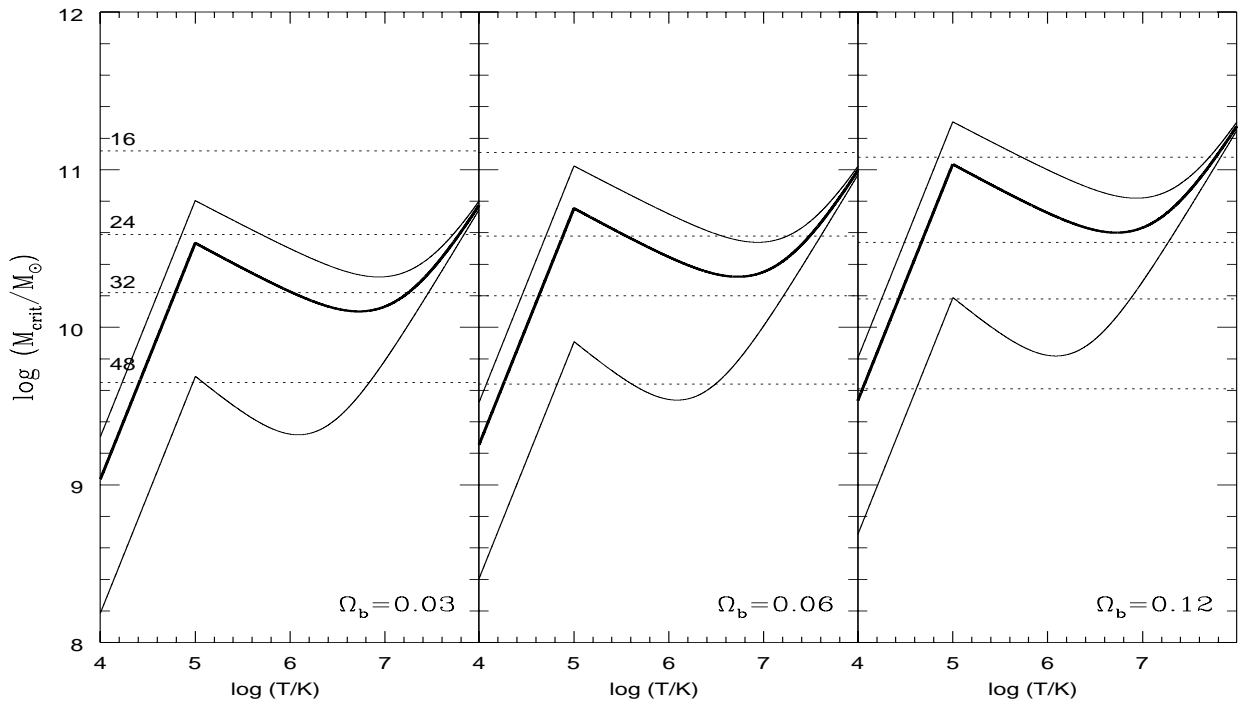,angle=270,height=10cm,width=18cm}}
\caption{The critical dark matter particle mass for spurious 2--body heating
to exceed cooling,
as a function of temperature (\protect\cite{SW}). The three solid curves in 
each plot correspond to different gas metallicities (from top to bottom,
$\Zmet/\Zsol=1.0,0.5,0.0$, with the fiducial metallicity, 0.5, in
bold). The different panels assume a local gas fraction, $f$, corresponding 
to the values of the global baryon fraction analysed in this paper,
$\Omegab=0.03,0.06,0.12$. The dotted horizontal lines give the
masses of the dark matter particles for $N=2\times X^3$, where $X$ is the
label on each line. A value of $5$ was assumed for the Coulomb logarithm.}
\label{fig:mcrit}
\end{figure*}

Comparison 9 consisted of changing the mass resolution by varying the total
number of particles. The initial power spectrum was truncated appropriately
to account for the varying Nyquist frequency of the particle mesh. The four
runs have $N=2\times(16,24,32,48)^{3}$ particles respectively, where the 2
indicates the number of species.  The masses of the two particle species in
each of these runs are given in Table~\ref{tab:pmass}. The three runs with
the largest $N$ were selected to quantify the result of increasing the
resolution of the simulation. We also added an $N=2\times16^3$ run to probe
the regime where artificial two--body heating of the gas particles by the
dark matter particles should significantly affect the cooling rate of the
gas. Figure~\ref{fig:mcrit} shows the critical dark matter particle mass
(defined as the mass where the cooling time of the gas equals the two-body
heating time) as a function of gas temperature,
for the metallicities and baryon fractions studied in this paper
(\cite{SW}). We have assumed the gas fraction to be equal to the
appropriate value of $\Omegab$, although the values can be scaled, if
desired. The Coulomb logarithm is assumed to be $5$, appropriate for
the largest haloes in the simulations studied in this paper (and therefore
demonstrating the worst case scenario for this effect). Overlaid as dotted
horizontal lines are the mass of the dark matter particles for the
simulations in this comparison. For $\Omega_b=0.06$ and $\Zmet=0.5 \Zsol$,
both the fiducial simulation and the $N=2\times48^3$ simulation are
acceptable for the range of temperatures relevant to this work, and the
$N=2\times24^3$ simulation is borderline. Since the $N=2\times16^3$ run
lies well above the critical line, the amount of gas cooled in this
simulation should be significantly affected by two--body heating.  The
effect is indeed severe: only 24 particles make it into the galaxy phase
and no galaxies form at all. (Note that since the cooling time at the
resolution limit of this simulation is $\sim 10^8$ years -- a small
fraction of the age of the universe -- in the absence of artificial heating
effects, we should have been able to form a population of cooled
objects.)

For the remaining three simulations, the amount of uncollapsed gas remains
roughly constant, and the biggest change is in the increasing amount of gas
that cools from the shocked phase to the galaxy phase with larger $N$. This is
reflected both in the galaxy fractions and in the number of galaxies
present in the final catalogues. This effect is expected when varying the
mass resolution of the simulations. When $N$ is increased, the sampling of
the density field improves and hence smaller mass haloes are able to be
resolved. The smallest haloes in the simulation will, on average, form
first, and therefore have higher density contrasts than the haloes that
form later. As discussed in Section~\ref{subsec:evolphase}, most of the
galaxy mass is accumulated in these objects.

\input tab_pmass

Finally, we investigate the effect of changing the
initial redshift of the simulation by considering runs with
$1+z_{i}=10,25$ and $50$ respectively (comparison 10).  The main
potential pitfall when choosing $z_i$ is that choosing too small a value
will suppress the amount of small--scale power, causing objects on
small scales to form too late. Hence the initial redshift should be
high enough that scales of the order of the initial softening should
not have already become non--linear (\ie $\delta \ll 1$). For the
$1+z_{i}=50$ run, the fiducial boundaries of the uncollapsed phase
are sufficient to segregate the uncollapsed gas from the rest, while
for the $1+z_{i}=10$ run, the temperature boundary 
is rescaled by a factor of 6.25 , \ie $(25/10)^{2}$. Evidently,
there are no significant changes in any of the gas phases 
or in the number of galaxies, when comparing all three runs.

\subsubsection{Physical Comparisons}

The mass fraction of baryons was varied in comparison 11, forming the set of
values $\Omegab=(0.03, 0.06,0.12)$. Varying the baryon fraction alters
the number density of ions and free electrons per simulation gas particle (and therefore
the masses of both the gas and dark matter particle), and consequently changes 
the cooling rate, since the emissivity scales as $n^2$. Again 
(see Figure~\ref{fig:mcrit}), it is important
that the baryon fraction be consistent with the critical mass of the
dark matter particles, given both the metallicity and the value of
$N$. Clearly for our choice of parameters, the value $\Omegab=0.03$ 
($Z = 0.5, N = 2\times32^3$) is borderline and this will contribute to
the reduction in the number of galaxies formed. Increasing the baryon fraction
has little effect upon the number of galaxies but they are
systematically heavier.

Comparison 12 was performed in order to examine the effect of changing the
metallicity of the gas. Each particle is assigned a constant amount of
{\it metals} (\ie of nuclei that have an atomic number
greater than 2). The metallicity affects the shape of the cooling function,
as illustrated in figure~\ref{fig:coolfn}. 
Also affected is the critical
mass of the dark matter particles (which depends on the cooling function) as shown
in Figure~\ref{fig:mcrit}. The simulation with $Z=0$ is clearly on
the wrong side of the line and the galaxy phase fractions are consequently affected:
slightly more material cools with the higher metallicity, but the
run with no metals shows a catastrophic drop in the amount of material
in the galaxy phase, the number of galaxies and the completeness.

Finally, in comparison 13, we created a set of runs with different
initial gas temperatures in order to measure
how the dynamical and thermal evolution of the gas depends on
its initial thermal state. We included two runs, one with
$T_i=0\kelvin$ and the other with $T_i=10^7\kelvin$. Although these parameter
choices may seem hard to justify in physical terms, we decided to stick with our
approach of varying a single parameter at a time. The fiducial boundaries of the 
uncollapsed phase are sufficient for the run with $T_i=0\kelvin$, however
the boundaries are redrawn for the run with $T_i=10^7\kelvin$, such that 
$s<2.5$ and $\rho/\left< \rho \right><50$, where 
$s={\rm log}_{10}(T/(\rho \,/\left<\rho\right>)^{2/3})$
measures the specific entropy of the gas. (Using these constraints makes it easier
to pick out the uncollapsed gas since it is isentropic.) For the 
run with $T_i=0\kelvin$, there are no significant differences in the measured quantities,
however the run with $T_i=10^7\kelvin$ shows a small increase in the uncollapsed gas
phase at the expense of the shocked gas. For this run, the gas cools as expected from
adiabatic expansion until $z \sim 5$. However by $z=3$ the uncollapsed phase is at 
temperatures below $10^4 \kelvin$, over an order of magnitude below the adiabatic temperature.
At these redshifts, the density and the cooling rate of the gas is high enough to 
allow it to radiate a significant fraction of its energy away. Consequently, only a small 
fraction of the gas is too hot to undergo gravitational collapse.

\subsection{The Mass Distribution of Galaxies}

\begin{figure*}
\centering
\centerline{\psfig{file=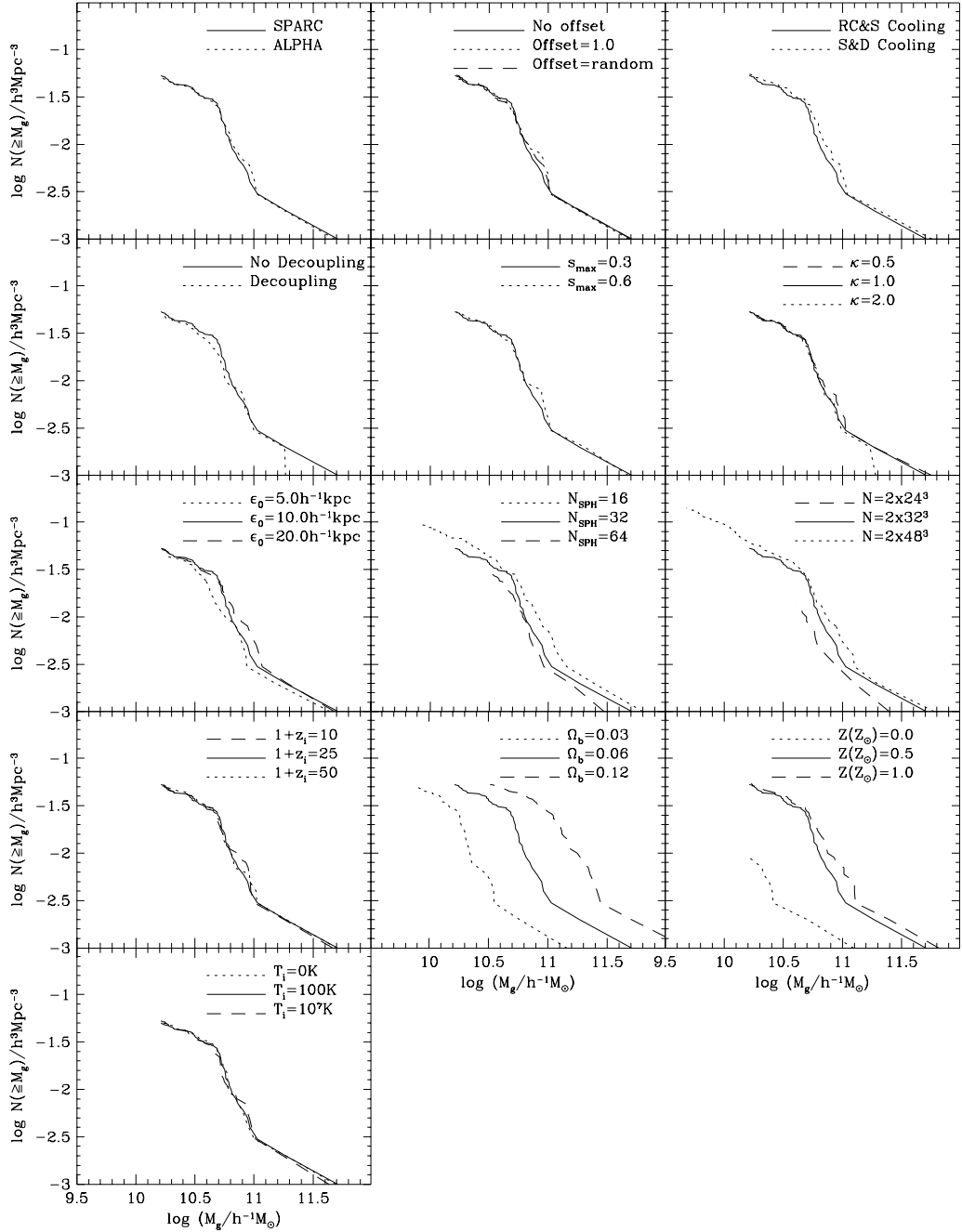,height=20cm}}
\caption{Cumulative galaxy mass functions for the full set of simulation 
comparisons. The plot gives the comoving number density of galaxies
with baryon mass greater or equal to $M_g$.
The fiducial simulation is always plotted with a solid line--type.}
\label{fig:cmf}
\end{figure*}

In Figure~\ref{fig:cmf} we plot the cumulative mass functions of
galaxies (\ie the number density of galaxies greater than a specified baryon
mass, $M_g$), with the fiducial simulation
always plotted as the solid line.  The dynamic range for these
simulations is about 2 orders of magnitude in abundance and
about 1.5 orders of magnitude in mass. The functions all have a
characteristic shape: a shallow slope at low masses, turning over to
a steeper slope at higher masses and a large tail at the high mass end. The
latter feature is simply due to the existence of a single massive galaxy that
has a baryonic mass of $\sim 5 \times 10^{11} \hMsol$. Galaxies this
massive are extremely rare in the universe; we shall discuss this
object below.

The biggest differences in the mass functions occur in the physical
comparisons, in which the gas cooling rate varies. 
The simulation with $\Zmet=0$ fails to produce as
many galaxies as the fiducial simulation -- the abundance is down by a
factor of 5 and the galaxies that do form are much less massive.
The large differences seen in the runs with varying baryon fraction
are mainly due to the mass of the individual gas particles being
different by a factor of two. However, when the galaxy masses are scaled 
to take out this effect, there is still a systematic difference 
between each function. This is due to the effect the baryon fraction
has on the cooling rate of the gas.

Varying the number of SPH neighbours shows a progression in the smallest
resolved galaxy mass due to the selection criterion that the number
of constituent particles $\Ng > \NSPH$. There are fewer galaxies for
larger $\NSPH$ and they are systematically less massive. This is a 
reflection of the fact that the algorithm has to smooth over a larger
range, which consequently leads to a poorer resolution of the density
field. Since emissivity scales as $n^2$, the lower cooling rate inhibits
the galaxy formation process - an entirely numerical effect. Varying the
value of $N$ shows similar results for the same reason, \ie a larger
number of particles enables the SPH algorithm to resolve higher gas
densities.

The remaining differences are relatively small. Changing the cooling 
function to the tabulated form of Sutherland \& Dopita (\shortcite{SUTDOP}) 
produces more cooling around $10^5\kelvin$, boosting the cooling rate 
and systematically increasing the mass of the galaxies by $\sim 10\%$. 
However, there is no significant difference in the shape of the mass
function.

Comparison 4 (in which the cold gas is decoupled from the hot gas)
also shows a small difference in the mass functions. 
This difference is dominated by the largest object in the 
simulation volume, which is reduced in mass by a factor of 3 when the cold,
dense gas is decoupled from the hot gas. Notably, this alteration to
the algorithm has a negligible effect on the smaller
objects.

The comparison set in which the final softening length is varied
shows an offset for $\sl0=5 \hkpc$. This simulation produces galaxies that
are systematically lighter than the simulations with larger values of $\sl0$,
as expected if they are being affected by two--body heating from the
dark matter particles.

\subsection{Comparison of Matched Galaxies and Haloes}

\input tab_scatterstats

To compare the individual properties of the galaxies and haloes, we
performed object--to--object matching between each simulation and the fiducial
run. Since the number of objects in each catalogue is different in
nearly every case, there is not a straightforward one--to--one
correspondence between the sets. To circumvent this, we always took the smallest
object set first and compared the larger one to it. When more than one object in 
the larger set was matched with the same object in the smaller set, the closest 
pair was selected. Generally, this was not a problem since the discarded
objects are usually at considerably larger distances than their matched object.
However, this method will inevitably produce outliers in both separation and mass 
differences due to the fact that, on occasion, an object in one simulation is
fragmented into smaller objects in another. This prompted us to use the median and
semi--interquartile ranges as statistical measures of the offset and scatter of any 
relative quantity we measured. Positions of the galaxies were defined as the median 
position of the linked particles and, for the haloes, their centre of mass. Below we 
detail the individual comparisons we made.

\subsubsection{Galaxy--galaxy displacements}

\begin{figure*}
\centering
\centerline{\psfig{file=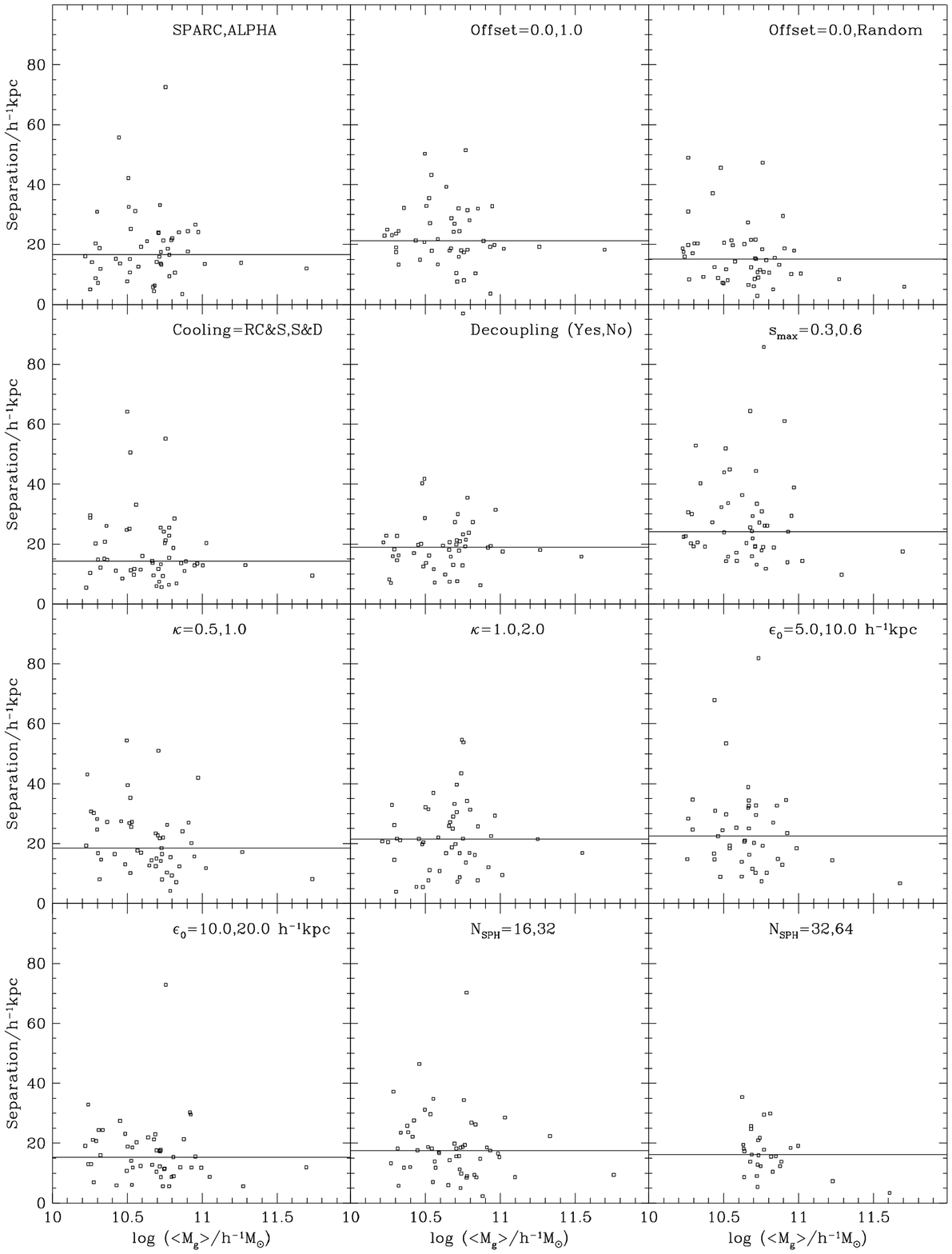,height=18cm}}
\caption{Separations between matched pairs of galaxies in the fiducial simulation
and all other simulations, plotted against the average galaxy mass.The labels 
indicate the parameter that varies in each plot. The median scatter is shown as a 
solid horizontal line. Matches above $100\hkpc$ are not plotted.}
\label{fig:pscat1}
\end{figure*}

\begin{figure*}
\centering
\centerline{\psfig{file=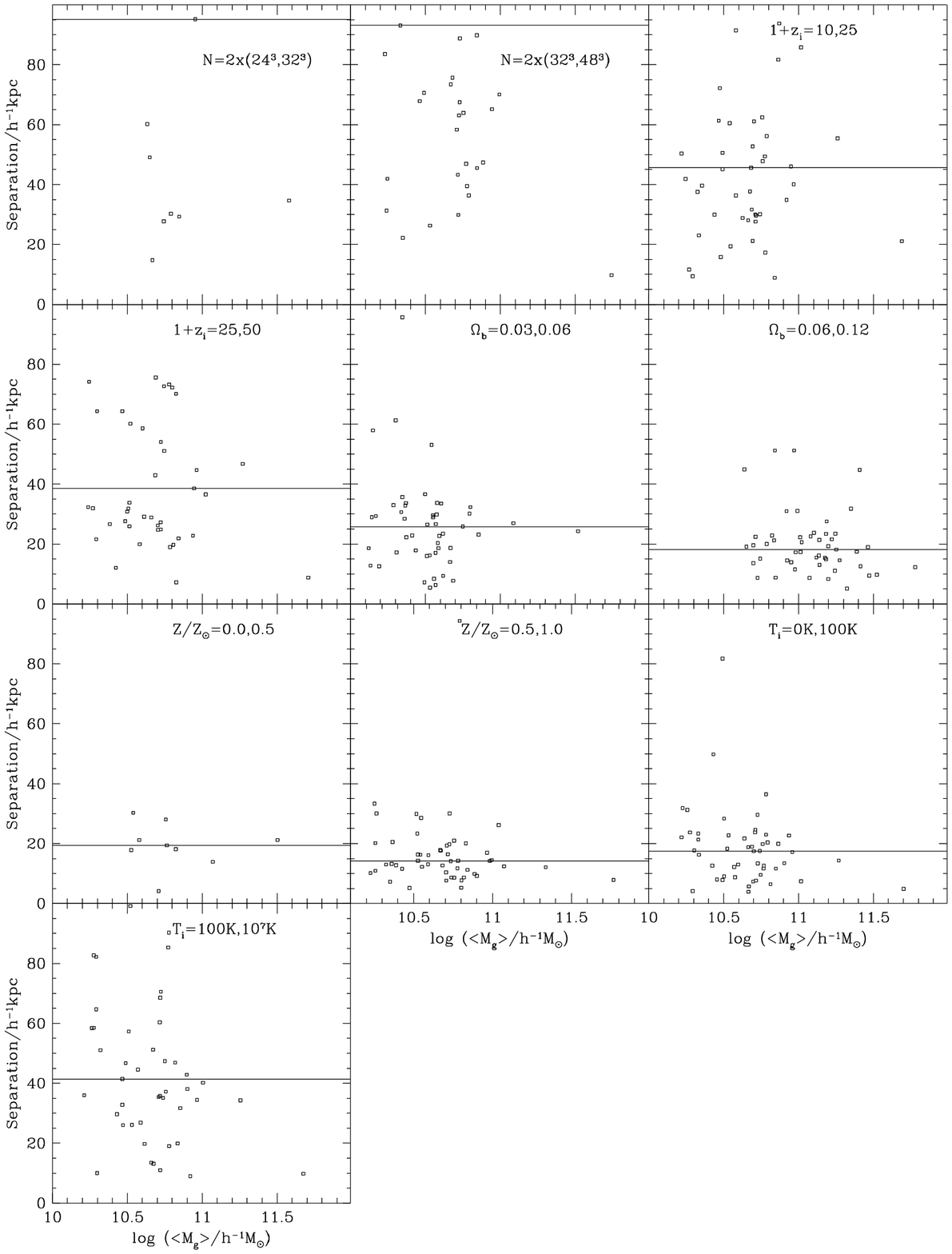,height=18cm}}
\caption{Separations between matched pairs of galaxies (continued).}
\label{fig:pscat2}
\end{figure*}

The first property we examined was the scatter in the positions of the
matched galaxies. The results are illustrated in Figures~\ref{fig:pscat1}
and \ref{fig:pscat2}, which show the displacement between each matched
pair (in $\hkpc$) plotted against the average object mass. The solid line
illustrates the median separation for each comparison. The number of
galaxies matched and the median plus semi--interquartile ranges of the
displacements are presented in Table~\ref{tab:scatterstats}, columns 3, 4
\& 5.  It is clear from the comparison between the simulations run on
different computer architectures that there is an intrinsic offset in
matched positions at the level of a few softening lengths. Furthermore, a
significant number of matches exceed this level by as much as a factor of
four. As a consequence of this, we regard as significant only those
comparisons that have median offsets in excess of the level seen for these
two simulations.

The first set to show a significant change is the simulations with
different values of $N$. Varying the mass resolution causes differences in
the number of objects formed and also affects the background dark matter
distribution, so it is of no surprise that the displacements show a
significantly larger amount of scatter in these comparisons. Varying the
value of $\NSPH$ also produces very different numbers of galaxies but this
does not affect the median separation at all, demonstrating that it is the
difference in the dark matter that primarily drives the increased scatter
in the galaxy positions rather than mergers or interactions between the
different numbers of galaxies.

A systematic difference in the matched positions is also introduced when
the initial redshift of the simulation is changed: the median displacement
increases by a factor of two in both comparisons. Variations in the initial
positions and velocities of the particles leads to asynchrony in the
subsequent spatial trajectories of the galaxies.

Regarding the physical comparisons, substantially increasing the initial 
temperature of the gas leads to a significant effect. Particles that have
been supplied with this much thermal energy naturally have greater pressure
support, which directly affects the equation of motion, and therefore the
particle trajectories.

\subsubsection{Galaxy--galaxy masses}

\begin{figure*}
\centering
\centerline{\psfig{file=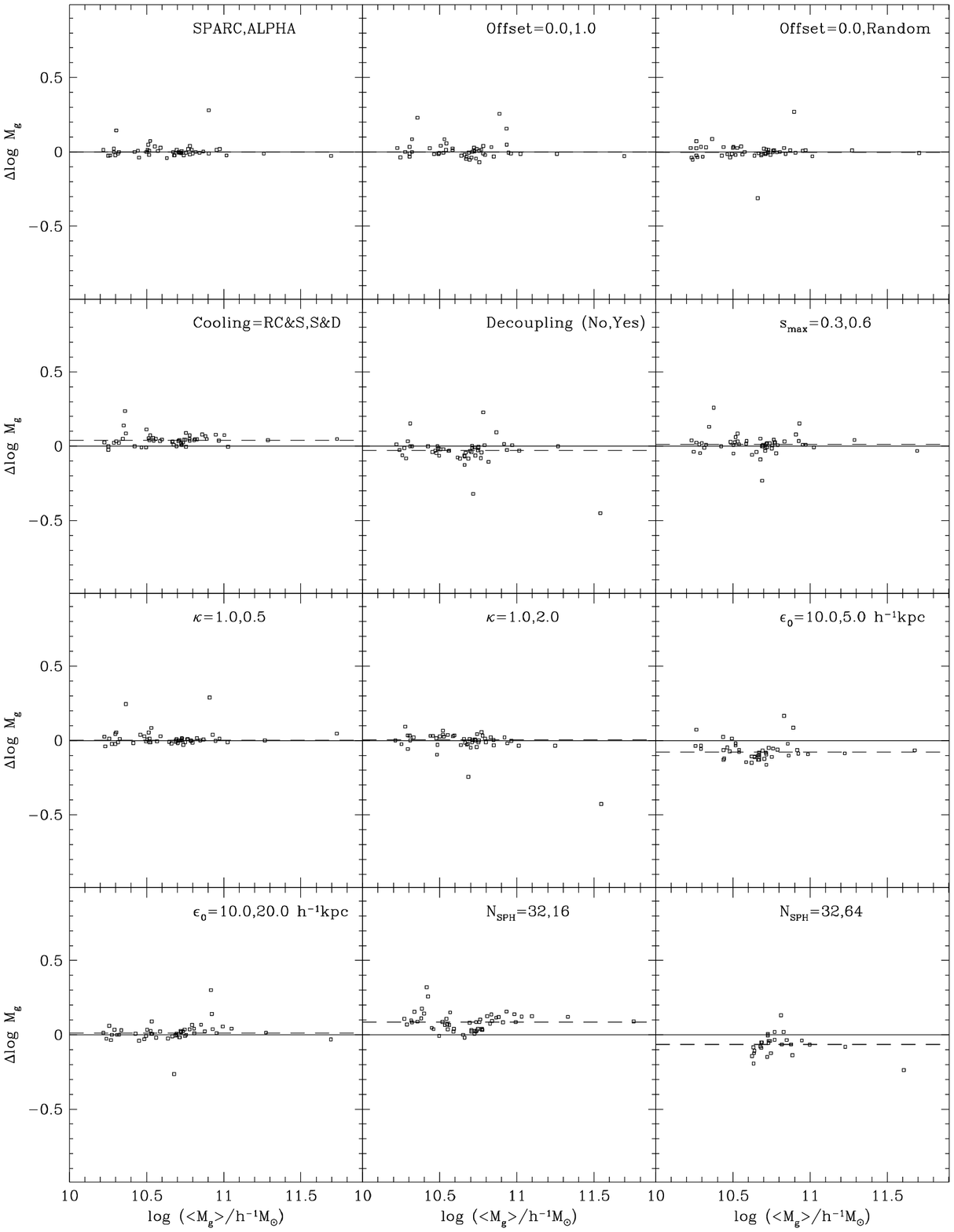,height=18cm}}
\caption{The logarithm of the ratio of the masses of matched pairs of galaxies
plotted against the logarithm of the mass of the fiducial galaxy. Labels indicate 
the parameter corresponding to each comparison (with the relative difference
defined as the second minus the first value). The solid line corresponds to equal
masses and the dashed line illustrates the median value for each comparison.}
\label{fig:mm1}
\end{figure*}

\begin{figure*}
\centering
\centerline{\psfig{file=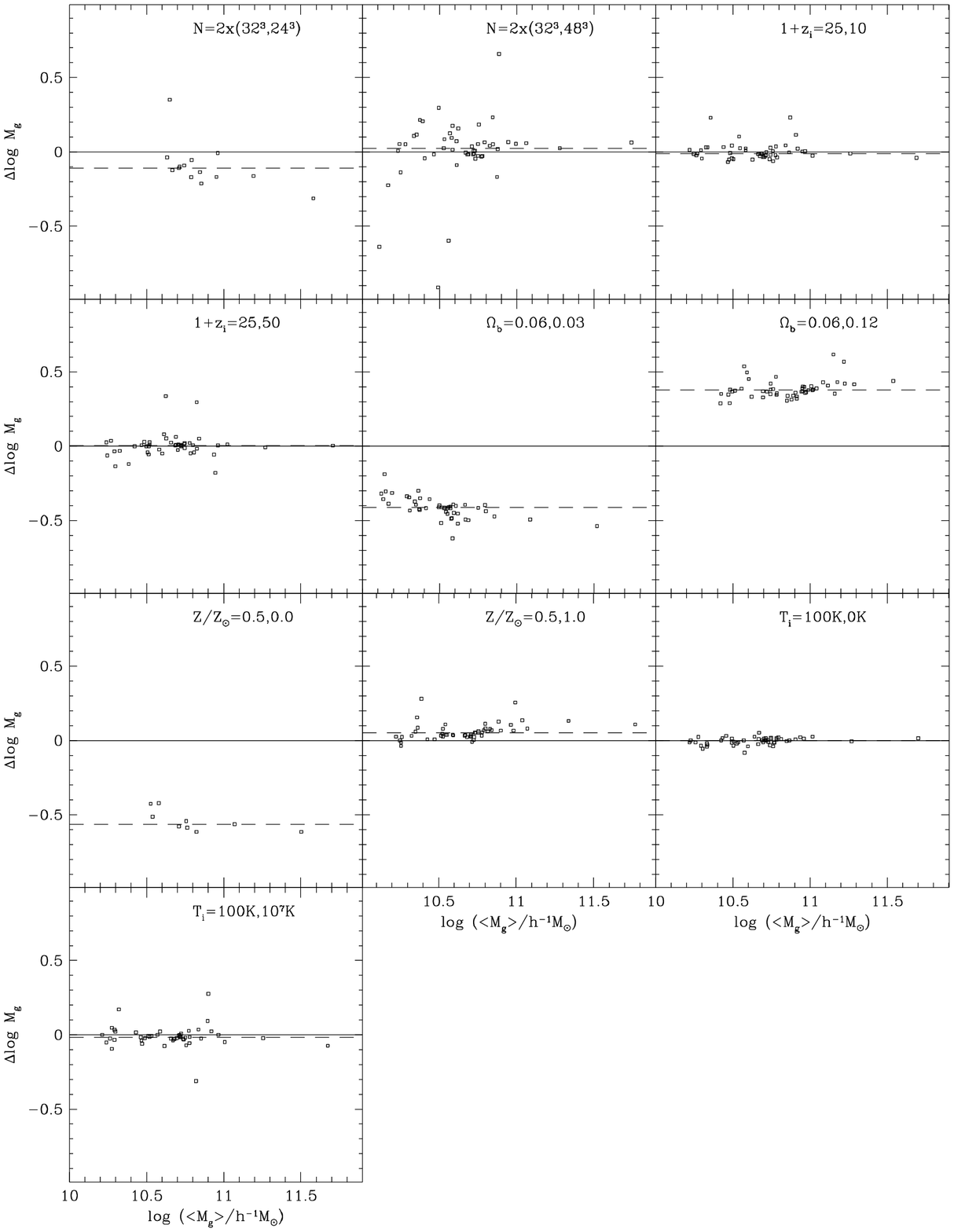,height=18cm}}
\caption{The logarithm of the ratio of the masses of matched pairs of galaxies (continued).}
\label{fig:mm2}
\end{figure*}

We have also analysed the scatter in the masses of the matched galaxy
pairs.  Figures~\ref{fig:mm1} and \ref{fig:mm2} illustrate the results by
showing the logarithm of the mass ratio of each matched pair plotted
against their average baryonic mass, always defined as the variant relative
to the fiducial case.  The solid line illustrates a ratio of unity (\ie
equal masses) and the dashed line illustrates the median value of the mass
ratio. The median and semi--interquartile ranges are given in columns 6 \&
7 of Table~\ref{tab:scatterstats}.

Adopting the tabulated cooling function causes an increase in the mass of
each galaxy, because of an overall enhancement in the cooling rate around
the temperatures most appropriate for these simulations ($\sim 10^5 - 10^7
\kelvin$).  Decoupling the hot gas from the cold gas causes the masses of
the galaxies to decrease slightly with only a few exceptions, notably the
largest object. Decreasing the softening length from $\sl0=10$ to $5 \hkpc$
also has the effect of decreasing the masses of the galaxies. Similarly, a
larger value of $\NSPH$ causes the masses of the galaxies to be
systematically lower, due to its effect on the density field and hence the
cooling rate.

The dispersion in the mass ratio for the comparisons with varying $N$ is
large because of the large scatter in object positions for these runs. This
makes spurious matches much more likely. For $N =2\times 24^3$ the objects
are significantly less massive, whilst for $N =2\times 48^3$ the objects
are only a little more massive than the fiducial simulation.

The largest differences in galaxy mass result from altering the cooling
rate via the physical parameters, $\Omegab$ and $\Zmet$.  For the
comparisons with varying baryon fraction, the galaxies are more than a
factor of two heavier for the larger value of $\Omegab$. Furthermore, there
appears to be a slight trend for increasing mass excess for heavier
galaxies but this is difficult to measure due to the small dynamic range in
mass. Varying the metallicity produces a similar effect to varying
$\Omegab$.

\subsubsection{Dark matter halo--halo masses}
The method used for comparing masses of matched galaxy pairs was also
performed on the dark matter haloes. Columns 8 \& 9 in
Table~\ref{tab:scatterstats} quantify the differences. Overall, very tight
correlations are seen for most of the simulations, as the dark matter is
the dominant source of the gravitational potential and so is not
significantly affected by changes in the gas dynamics. The simulation with
$\Omega_b=0.12$ shows the largest difference -- here the contribution 
from the gas starts to make a difference to the properties of
the dark matter haloes.

\subsection{The Distribution of Baryons in Haloes}

The final item we investigate in this study is how parameter variations affect
the final distribution of the baryons within haloes, as a function
of the halo virial mass. Specifically, we looked at the mass fraction of baryons
present in the galaxy phase, the ratio of baryon masses for matched haloes 
and the total halo baryon fraction normalised to the global value.

\subsubsection{Fraction of baryons in galaxies}

\begin{figure*}
\centering
\centerline{\psfig{file=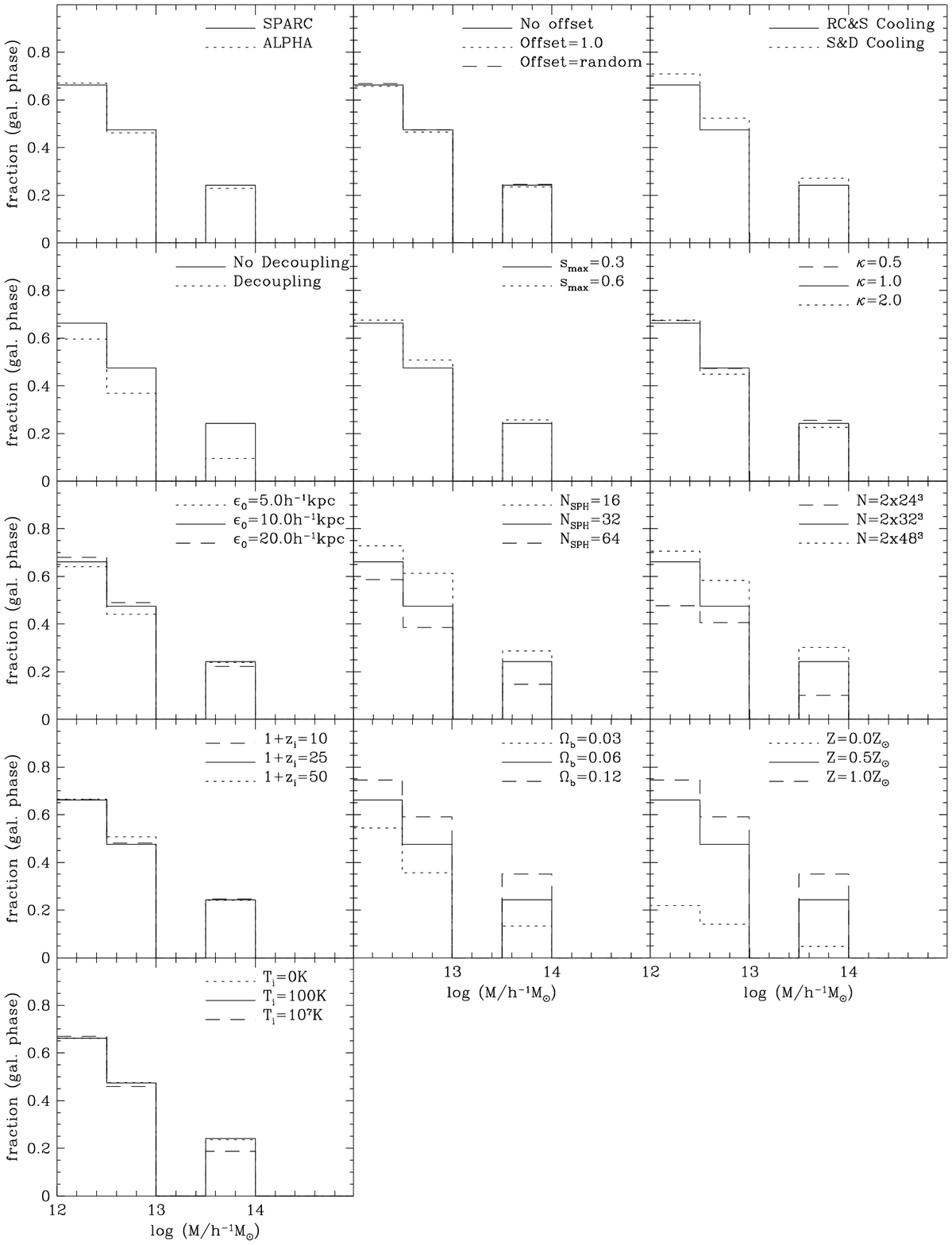,height=20cm}}
\caption{The mass fraction of baryons in the galaxy phase plotted as a 
function of the total halo mass. The fiducial simulation is the solid 
histogram.}
\label{fig:frac}
\end{figure*}

Figure~\ref{fig:frac} illustrates the fraction of baryons in the galaxy
phase as a function of the halo virial mass, for each comparison. Because
of our small samples, we have binned the data as histograms; the last bin
corresponds to the single largest object. This statistic is remarkably
stable; the first two comparisons show virtually no difference. The trend
is for higher mass haloes to have less of their mass in galaxies. The mass
fraction varies from $\sim 0.65$ in the smallest haloes down to $\sim 0.2$
in the highest mass halo. The comparison between the fiducial simulation
and the case in which the galaxy phase is decoupled from the hot gas
reflects the reduction in the total amount of galaxy phase baryons in the
latter -- about a factor of three for the galaxy in the largest halo ($\sim
5\times 10^{13} \hMsol$), but only $\sim 10\%$ for haloes with masses of
$\sim 10^{11}\hMsol$. Other comparisons that give significant changes in
the mass fractions show nothing unexpected.

\subsubsection{Local halo baryon fraction}

\begin{figure*}
\centering
\centerline{\psfig{file=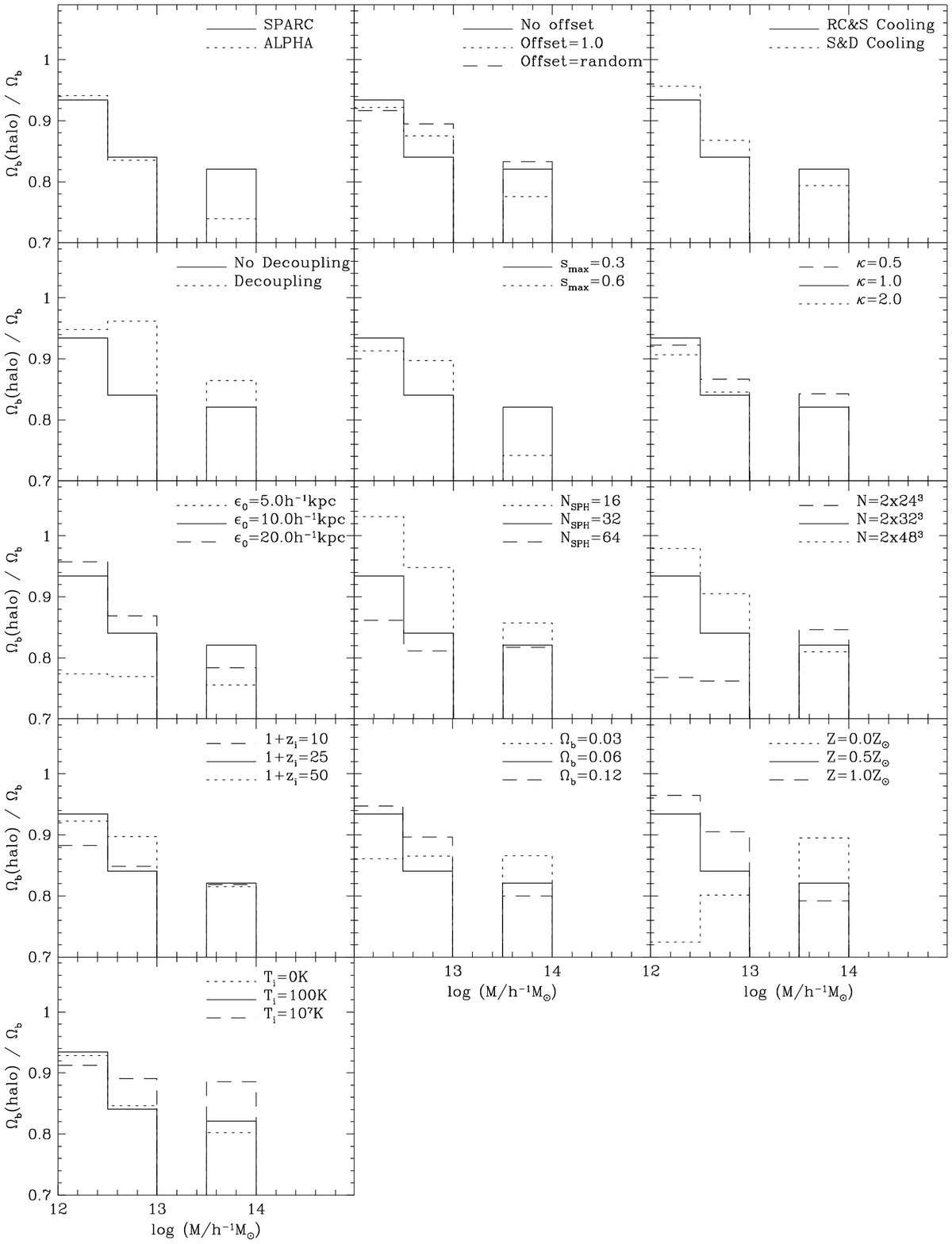,height=20cm}}
\caption{The baryon fraction in haloes as a function of the total halo mass, 
in units of the global baryon fraction. The fiducial simulation is the solid 
histogram.}
\label{fig:bfrac}
\end{figure*}

The variations in the total mass of baryons within the virial radius of
matched haloes are quantified in columns 10 \& 11 of
Table~\ref{tab:scatterstats}.  The median differences are slightly higher
than for the dark matter, but smaller than the galaxies
alone. Interestingly, the sign of the median difference is not always the
same as the equivalent comparison for the galaxies, indicating that
fluctuations in the total baryon content are not due to the galaxies
alone. Significant differences are most evident in the simulations in which
parameter changes have affected the cooling rate. Inducing a higher cooling
rate causes the gas to lose some of its pressure support and therefore to
become more compressed within its halo. The simulation with $\sl0=5\hkpc$
shows a significant change in the same direction as the change in the
galaxy masses, supporting the idea that the gas is being more strongly
heated from two--body encounters with the dark matter particles.

We also analysed the total baryon fraction in haloes as a function of the
halo virial mass. The results are displayed in Figure~\ref{fig:bfrac}.  The
fraction is normalised to the global value of $\Omegab$ for each
simulation. The fiducial simulation has halo baryon fractions ranging from
$\sim 93\%$ of the global value in the smallest haloes to $\sim 82\%$ in
the largest. All simulations agree with the fiducial values to within $\sim
20\%$. Notably, not all simulations show a monotonically
decreasing baryon fraction with halo mass. However, the scatter is 
considerable because of the small number of objects in each bin (with
only one in the highest mass bin). The differences reflect the varying
amounts of material that has cooled. Note, for example, the drop in the
first bin for $\Zmet=0.0$, for which galaxies do not form in small haloes
and the rise in the first bin for $\NSPH=16$. For almost all of the
simulations and across the entire range of halo masses, the baryon fraction
is less than the global value.  The baryon fraction in the
largest halo in this simulation (0.82 for the fiducial case) is
somewhat smaller than the values, 0.9-1.0, found in the rich galaxy cluster
simulations presented by Frenk \etal (1999). The value for our largest
cluster, however, is noisy as indicated by the first comparison in
the figure and it depends on metallicity, as indicated also in the
figure. Furthermore, this cluster is approximately 10 times less massive
than the cluster in Frenk \etal and, as shown by Pearce \etal (in
preparation), there is a weak trend of increasing baryon fraction with mass
for clusters in the SCDM cosmology. 

\section{Summary of the comparison results}
\label{sec:summary}

We now summarise the main effects seen in each comparison, in turn.

In comparison 1, we looked at the effects introduced by running the
simulation on different computer architectures.  No significant differences
were found between the two simulations. However, the displacements between
matched pairs of galaxies at $z=0$ were found to be of the order of an S2
softening length. It is therefore incorrect to trust either the positions
of individual galaxies or their trajectories on scales smaller than this.

Comparison 2 was intended to look at the effects of small displacements in
the positions of the particles at the start of the simulation. We ran one
simulation with the particles initially displaced by 1 FFT cell to test the
periodic nature of the algorithm. We also ran a simulation with tiny random
perturbations applied to the initial particle positions, to test the level
at which differences can grow over the course of a few thousand timesteps
(the typical length of each simulation). Again, no significant changes were
detected amongst the three simulations.

We next changed the cooling function from the series of power--law fits,
used in the fiducial simulation, to a set of tabulated values from
Sutherland \& Dopita (\shortcite{SUTDOP}). A slight enhancement in the galaxy masses,
at the $10\%$ level, was produced in the latter case, although no significant changes 
were produced in the spatial distribution of galaxies.

In comparison 4, we compared the fiducial simulation to one in which the
galaxy phase gas is partially decoupled from hot gas in the SPH algorithm (see
\cite{VGF1}). This modification provides a better estimate of the hot gas density 
near galaxies, specifically it prevents an overestimate which leads to an artificial 
enhancement in the cooling rate. Only the largest galaxy in the  simulation was 
significantly affected, with its mass being reduced by a factor of three.

Comparison 5 consisted of doubling the initial (maximum) softening length.
This results in a period of time in which the softening is larger than in the
fiducial simulation. This led to  slight differences in the spatial
distribution of the galaxies, although it left the mass distribution unaffected.

We altered the size of the timesteps by a factor of two each way, in
comparison 6, by changing the constant, $\dtnorm$. Increasing the
timestep resulted in much more hot gas at the expense of the cold phase. This
illustrates the effect of the timestep on the dynamics of the particles: 
introducing significant errors into the positions and velocities leads to 
enhanced heating of the gas. The galaxies themselves remained largely unaffected.
No significant change was evident when halving the value of $\dtnorm$, thus
justifying the value $\dtnorm=1$, adopted as fiducial.

Comparison 7 consisted of changing the final value of the softening length, $\sl0$. 
In the simulation with $\sl0=5\hkpc$, the galaxies were about $20\%$ less massive than 
in the fiducial simulation ($\sl0=10\hkpc$), possibly due to discreteness effects caused
by harder two--body encounters. A larger value of $\sl0=20\hkpc$, which implies a 
constant comoving softening, produced no significant differences.

In comparison 8, we varied the value of $\NSPH$, making it larger and smaller 
by a factor of two, thus looking at the effects produced when the SPH
algorithm uses a different number of neighbours when calculating the smoothing
length for each gas particle. These changes had a dramatic effect on the simulations.
The value of $\NSPH$ determines the minimum mass for which the gas density can be resolved:
smaller values will resolve smaller masses at the expense of larger random 
fluctuations. We found that the number of galaxies increases strongly with smaller
$\NSPH$, as did the masses of the matched galaxy pairs, although without significantly 
affecting their positions.

Comparison 9 consisted of changing the number of particles in the simulation,
thus altering the mass resolution. To demonstrate the effect of spurious heating 
by dark matter particles, we ran a simulation with a dark matter particle mass
above the critical value calculated by Steinmetz \& White (\shortcite{SW}). In
this case, gas is unable to cool and form galaxies. For the main set of
comparisons, we had a suite of runs with $N = 2\times[24,32,48]^{3}$. 
The number and mass of galaxies increased with $N$. Improving the mass resolution
also affected the final distribution of dark matter. The two simulations with
the highest resolution were in better agreement than the two with the lowest resolution,
albeit that the discrepancies in the galaxy phase material (\eg the 
fraction of cooled gas) were still large.

In comparison 10, the initial redshift of the simulations was
changed ($1+z_i=10,25,50$). The masses of the objects were not significantly 
altered by this change but small changes in the spatial distribution of the galaxies
were produced.

Comparison 11 involved changing the value of the baryon fraction
($\Omegab$=0.03,0.06,0.12). The higher value of $\Omegab$ causes the
cooling rate to increase (since emissivity is proportional to
$\Omegab^2$). The mass of each galaxy also increases, since $m$ is
proportional to $\Omega_b$. The run with $\Omegab=0.03$ has dark
matter masses of the order of the critical mass for 2--body heating, 
although the reduction in the cooling rate produced by this effect is confused by 
the fact that the cooling rate is significantly less anyway. Overall, large 
differences were noted in all analyses in which the baryon fraction was varied. 

In comparison 12, we changed the value of the (fixed) gas metallicity 
($\Zmet=0.0,0.5,1.0$ in solar units). The run with no metals differed dramatically
from the fiducial simulation, producing very few objects by $z=0$. This simulation 
had a dark matter mass well above the critical value for 2--body heating. The run 
with $\Zmet=1.0\Zsol$ showed an enhancement in galaxy masses relative to the fiducial 
simulation, although the change was far less severe.

Finally, in comparison 13 the initial gas temperature was changed from $100
\kelvin$ to $0 \kelvin$ and $10^7 \kelvin$ respectively. No significant
changes were measured in the former two. Although the uncollapsed gas in the latter 
simulation is on a higher adiabat ($T \sim 10^3 \kelvin$),
this had very little effect on the final state of the gas in galaxies and haloes,
with the exception of introducing significant displacements between matched galaxy pairs.

\section{Conclusions}
\label{sec:conc}

In this paper, we have explored how variations in simulation parameters and
physical assumptions affect the population of galaxies which form in
cosmological simulations. We used the $\AP3M$ algorithm to calculate the
gravitational potential field and SPH with radiative cooling to model 
gas dynamics. We started with a carefully chosen fiducial simulation and
calculated a number of variants, each differing in only one
parameter. Galaxy and halo catalogues were selected in a consistent manner
before analysing the effects that the parameter changes had on the
thermodynamic state of the gas, on the mass and spatial distribution of the
galaxies, and on the baryon content of haloes.

By carefully tracking the phase evolution of gas particles we have shown
that most galaxy material (around $90\%$ in the fiducial simulation)
originally cools in small haloes where the virial temperature is $\sim
10^5\kelvin$. As Figure~\ref{fig:frac} shows, the percentage of halo
baryons that are cold rises as the halo mass falls. Only a small fraction
of the baryons cool at later stages of the evolution. Mergers of these
small fragments can reheat the gas briefly as massive galaxies are built
up. This implies that the most dynamically interesting part of the process
occurs at the resolution threshold of the simulation and is therefore
sensitive to many numerical and physical parameters.

Numerical noise leads to a scatter in the final galaxy positions that is of
the order of the (spline) softening length for most galaxies, but can be
much higher in some cases. Individual galaxy trajectories should therefore
be treated with caution.

The length of timestep recommended by Couchman, Thomas \& Pearce
(\shortcite{CTP}) is adequate. Halving it caused no significant change in
the galaxy catalogues, but choosing a larger value led to a significant
increase in the shocked gas relative to the uncollapsed gas. This can be
attributed to errors in the particle positions and velocities, resulting in an
artificial enhancement in the thermal heating of the gas.

We confirm that simulations with dark matter particle masses greater than
the Steinmetz \& White (\shortcite{SW}) critical mass for 2--body heating
do indeed show a catastrophic reduction in the final amount of cooled gas
and in the number of galaxies, even when the shortest cooling time is a
small fraction of the age of the universe. The choice of dark matter
particle mass should be comfortably below this critical value. The mass in
galaxies is also strongly affected by the choice of baryon fraction and
metallicity.

Pre--heating of the gas to a high initial temperature ($10^{7}\kelvin$ at
$z=24$) produces small changes in the final positions of the galaxies 
but has little effect on their final mass. A significant fraction of the
thermal energy of the gas is radiated away around $z=3-5$, allowing the
majority of the baryons to undergo gravitational collapse.

Increasing the mass resolution by using a larger number of particles causes
both the number of galaxies and their masses to increase. Improved
resolution of the density field enhances the cooling rate within haloes,
particularly at high redshift when the first objects form. As the mass
resolution of the simulation increases, modelling feedback processes
correctly becomes critical in order to avoid overcooling of baryonic
material.

Lowering the number of neighbours used by the SPH algorithm to calculate
hydrodynamical quantities has a similar effect on the amount of cooled
material to increasing the total number of particles in the simulation.
Smaller values of $\NSPH$ enable higher overdensities to be resolved
(therefore enhancing the cooling rate in such regions), at the expense of
more contamination from sampling noise.

The fraction of the baryonic material that cools within a halo and forms
galaxies is a very stable quantity which varies inversely with halo mass. 
Reducing the total amount of gas affects all scales in a similar way. The 
ratio of baryonic to dark material within the virial radius of haloes is 
found to be around $0.8 \pm 0.1$ of the global value, for the range of 
halo masses studied here. 
 
In this study we have used four different prescriptions for the
gravitational softening. We have varied the initial comoving softening,
from $20\hkpc$ to $40\hkpc$, and we have also used a range of final
physical softenings, from $5\hkpc$ to $10\hkpc$ and $20\hkpc$. With the
exception of the $5\hkpc$ case, this variation had very little effect on
the galaxy population.  The smaller final softening systematically reduced
both the mass of galaxies and the mass of baryons within haloes by $\sim
20\%$. This effect is probably due to the artificial heating of the gas
from two--body encounters with the dark matter particles discussed above
(\cite{SW}). The fiducial softening is borderline for this resolution.

Finally, in almost every simulation we obtained an overly massive galaxy at
the centre of the largest halo. With a baryonic mass of $\sim 5 \times
10^{11} \hMsol$, such an object is not expected within such a small volume
of the universe. Decoupling the central dense gas from the hot halo
material effectively reduces the mass of this object to a more reasonable
value (see also \cite{VGF1}).

We conclude that varying most of the numerical or physical parameters
required in simulations of galaxy formation within reasonable bounds
produces only relatively small changes in the abundance, mass and spatial
distribution of the galaxies that form. The most important numerical
parameter is mass resolution which has an important effect because most of
the gas cools in small mass haloes (in which the cooling times are
shortest). The most important physical parameters are those that control
the cooling time of the gas, the density of baryons and its
metallicity. Varying these within plausible ranges can substantially
change the abundance and mass of galaxies. Careful modelling of physical
processes occurring in small mass haloes where the thermodynamic state of the
gas is likely to be affected, for example, by energy injected by supernovae
and stellar winds is the next great challenge facing simulators of galaxy
formation.


\section*{Acknowledgements}
We would like to thank the referee Matthias Steinmetz for his 
constructive suggestions that have improved the clarity of the manuscript.
The work presented in this paper was carried out as part of the programme
of the Virgo Supercomputing Consortium
(http://star--www.dur.ac.uk/$\sim$frazerp/virgo/). We acknowledge
a NATO collaborative research grant (CRG 970081). This work was supported
by the EC network for ``Galaxy formation and evolution''.  STK acknowledges
the support of a PPARC postgraduate studentship; CSF acknowledges a PPARC
Senior Research Fellowship and a Leverhulme Research Fellowship.

\end{document}

%% file: tab_fidparam.tex
\begin{table}
\caption{The values of the parameters chosen for the fiducial simulation that
were subsequently varied. $N$ is the total number of particles, $\NSPH$ is the
number of gas particles over which the SPH smoothing is carried out, $L$
is the size of the FFT mesh, $\slmax$ is the maximum comoving size
of the softening in units of FFT cells, $\sl0$ is the effective Plummer
softening, $\dtnorm$ is the timestep normalisation, $\Omegab$ is
the baryon fraction, $\Zmet$ is the gas metallicity, $z_{i}$ is the
initial redshift and $T_{i}$ is the initial gas temperature. The choice
of these parameters is discussed in more detail in the text.}
\begin{center}
\begin{tabular}{ll} 
Parameter & Value\\
$N$		&	$2\times32^{3}$\\
$\NSPH$		&	$32$\\
$L$		&	$64$\\
$\slmax$	&	$0.3$\\
$\sl0$		&	$10 \, \hkpc$\\
$\dtnorm$	&	$1$\\
$\Omegab$	&	$0.06$\\
$\Zmet$		&	$0.5 \Zsol$\\	
$z_{i}$	 	&	$24$\\
$T_{i}$		&	$100\kelvin$\\
\end{tabular}
\label{tab:fidparam}
\end{center}
\end{table}

%% file: tab_select2.tex
\begin{table*}
\caption{Details of the set of simulations analysed in this paper, following the 
various definitions and selection
procedures discussed in Section \ref{sec:fiducial}. The first
column is the comparison number given to the simulation, 
with the second column indicating the alteration made to the
fiducial initial conditions. The next three columns give the fraction
of baryons in each of the three phases defined in the text. Columns six
and seven give the number of galaxies and haloes respectively that
remain in the catalogue after the various selection procedures have
been applied. The
final column is the {\it completeness}, i.e. the fraction of baryons
in the galaxy phase that are present in the final galaxy
catalogues. Bold type has been used to denote values that deviate
significantly from the fiducial case (two per cent for the phase fractions,
five percent for the galaxies, haloes and completeness).}
\begin{tabular}{llllllll} 

No. & Alteration & Uncollapsed Phase & Shocked Phase & Galaxy Phase & Galaxies & Haloes & Completeness\\

0       & Fiducial Simulation           & 0.18          & 0.64          & 0.18          & 53 & 31 & 0.96\\

 1      & Architecture                  & 0.16          & 0.65          & 0.19          & 51 & 32 & 0.96\\
 2      & Position Offset (1 cell)      & 0.17          & 0.65          & 0.18          & 50 & 31 & 0.96\\
 2      & Position Offset (random)      & 0.18          & 0.64          & 0.18          & 55 & 30 & 0.97\\
 3      & New Cooling Function          & 0.18          & 0.62          & 0.20          & 55 & 31 & 0.98\\
 4      & Cold gas decoupled            & 0.18          & {\bf 0.67}    	& {\bf 0.15}	& 51 & 31 & 0.95\\

 5      & $\slmax=0.6$                  & 0.17          & 0.64          & 0.19          & 53 & 30 & 0.96\\
 6      & $\dtnorm=0.5$                 & 0.18          & 0.63          & 0.19          & 52 & 31 & 0.97\\
 6      & $\dtnorm=2.0$                 & {\bf 0.06}    & {\bf 0.78}    	& 0.18          & 53 & 31 & 0.95\\
 7      & $\sl0=5\hkpc$                 & 0.16          & {\bf 0.69}    & {\bf 0.15}          & {\bf 45} & 31 & 0.95\\
 7      & $\sl0=20\hkpc$                & 0.18          & 0.63          & 0.19          & 53 & 30 & 0.97\\
 8      & $\NSPH=16$                    & 0.18          & {\bf 0.56}    & {\bf 0.26}    & {\bf 97} & 31 & 0.98\\
 8      & $\NSPH=64$                    & 0.17          & {\bf 0.70}    & {\bf 0.13}    & {\bf 28} & 31 & {\bf 0.89}\\
 9      & $N=2\times16^3$               & 0.20          & {\bf 0.80}    & {\bf 0}       & {\bf 0 } & {\bf 29} & 0\\
 9      & $N=2\times24^3$               & 0.17          & {\bf 0.74}    & {\bf 0.09}    & {\bf 14} & 30 & {\bf 0.71}\\
 9      & $N=2\times48^3$               & 0.17          & {\bf 0.57}    & {\bf 0.26}    & {\bf 141}& {\bf 34} & 0.98\\
10      & $1+z_{\rm i}=10$              & 0.17          & 0.65          & 0.18          & 55 & 32 & 0.97\\
10      & $1+z_{\rm i}=50$              & 0.19          & 0.63          & 0.18          & 51 & {\bf 29} & 0.96\\

11      & $\Omegab=0.03$                & 0.18          & {\bf 0.69}    & {\bf 0.13}    & {\bf 49} & 31 & 0.94\\
11      & $\Omegab=0.12$                & 0.18          & {\bf 0.58}    & {\bf 0.24}    & 53 & 31 & 0.97\\
12      & $\Zmet=0.0\Zsol$              & 0.18          & {\bf 0.78}    & {\bf 0.04}    & {\bf 9}  & 30 & {\bf 0.44}\\
12      & $\Zmet=1.0\Zsol$              & 0.18          & {\bf 0.60}    & {\bf 0.22}    & 54 & 31 & 0.97\\
13      & $T_{\rm i}=0\kelvin$          & 0.18          & 0.64          & 0.18          & 51 & 31 & 0.95\\
13      & $T_{\rm i}=10^7\kelvin$       & {\bf 0.22}     & {\bf 0.61}    & 0.17          & 50 & 30 & 0.96\\

\end{tabular}
\label{tab:select}
\end{table*}

%% file: tab_pmass.tex
\begin{table}
\caption{The masses of each particle species for the runs performed
with varying values of $N$.}
\begin{center}
\begin{tabular}{lcc} 
$N$ & $\Mdark (10^{10} \hMsol)$ & $\Mgas (10^{10} \hMsol)$\\

$2 \times 16^{3}$ & 6.38 & 0.41\\
$2 \times 24^{3}$ & 1.89 & 0.12\\
$2 \times 32^{3}$ & 0.80 & 0.05\\
$2 \times 48^{3}$ & 0.24 & 0.015\\

\end{tabular}
\end{center}
\label{tab:pmass}
\end{table}

%% file: tab_scatterstats.tex
\begin{table*}
\caption{Statistical measures of the scatter present in all simulation
comparisons of the matched galaxies and haloes. The first two columns
detail the particular comparison and the third column lists the number
of galaxies in each catalogue that were matched with the fiducial simulation,
using the method detailed in the text. The rest of the columns list the
median and semi-interquartile range (siqr) for the measured scatter in galaxy
separation (in $\hkpc$), galaxy mass, halo dark matter mass and halo
baryon mass respectively.}
\begin{center}
\begin{tabular}{lllllllllll} 
No. 	& Alteration 			& Matches
						& \multicolumn{2}{c}{$\Delta r$(galaxies)} 	
						& \multicolumn{2}{c}{$\Delta \log M$(galaxies)}
				 		& \multicolumn{2}{c}{$\Delta \log M$(dark matter)}
						& \multicolumn{2}{c}{$\Delta \log M$(baryons)}\\
	&&					& median & siqr	& median & siqr	 & median & siqr  & median & siqr\\

1	& Architecture 			& 50	& 17. & 6.0  &  $\phantom{-}0.00$  & 0.01 & $\phantom{-}0.003$  & 0.01 & $-0.002$            & 0.01\\
2	& Position Offset (1 cell)	& 49	& 21. & 5.4  &  $\phantom{-}0.00$  & 0.02 & $\phantom{-}0.007$  & 0.02 & $\phantom{-}0.0$    & 0.02\\
2	& Position Offset (random) 	& 52	& 15. & 5.5  &  $\phantom{-}0.00$  & 0.02 & $\phantom{-}0.004$  & 0.009& $\phantom{-}0.002$  & 0.01\\
3	& New Cooling Function 		& 53	& 14. & 5.8  &  $\phantom{-}0.04$  & 0.01 & $\phantom{-}0.0$    & 0.01 & $\phantom{-}0.01$   & 0.02\\
4	& Cold gas decoupled		& 49	& 19. & 4.0  &  $-0.03$            & 0.03 & $-0.008$            & 0.01 & $-0.02$             & 0.03\\

5	& $\slmax=0.6$ 			& 51	& 24. & 7.3  &  $\phantom{-}0.01$  & 0.02 & $\phantom{-}0.002$  & 0.02 & $\phantom{-}0.02$   & 0.02\\
6	& $\dtnorm=0.5$ 		& 51	& 19. & 7.1  &  $\phantom{-}0.00$  & 0.01 & $-0.002$            & 0.01 & $\phantom{-}0.004$  & 0.01\\
6	& $\dtnorm=2.0$ 		& 50	& 22. & 7.4  &  $\phantom{-}0.00$  & 0.03 & $\phantom{-}0.004$  & 0.01 & $-0.002$            & 0.02\\
7	& $\sl0=5\hkpc$ 		& 44	& 23. & 8.7  &  $-0.08$            & 0.03 & $-0.007$            & 0.02 & $-0.07$             & 0.02\\
7	& $\sl0=20\hkpc$ 		& 50	& 15. & 4.9  &  $\phantom{-}0.01$  & 0.02 & $\phantom{-}0.0$    & 0.01 & $\phantom{-}0.01$   & 0.01\\
8	& $\NSPH=16$ 			& 53	& 18. & 5.9  &  $\phantom{-}0.09$  & 0.04 & $-0.003$            & 0.02 & $-0.04$             & 0.02\\
8	& $\NSPH=64$		 	& 28	& 16. & 4.3  &  $-0.06$            & 0.04 & $-0.003$            & 0.02 & $-0.02$             & 0.03\\
9	& $N=2 \times 24^3$ 		& 14	& 95. & 62.  &  $-0.13$            & 0.06 & $-0.01$             & 0.05 & $\phantom{-}0.07$   & 0.05\\
9	& $N=2 \times 48^3$		& 51	& 93. & 55.  &  $\phantom{-}0.02$  & 0.06 & $\phantom{-}0.005$  & 0.03 & $\phantom{-}0.01$   & 0.05\\
10	& $1+z_{\rm i}=10$ 		& 50	& 46. & 16.  &  $\phantom{-}0.01$  & 0.03 & $\phantom{-}0.0$    & 0.04 & $\phantom{-}0.01$   & 0.03\\
10	& $1+z_{\rm i}=50$ 		& 48	& 39. & 23.  &  $\phantom{-}0.00$  & 0.03 & $-0.001$            & 0.03 & $-0.004$            & 0.03\\

11	& $\Omegab=0.03$		& 48	& 26. & 7.5  &  $-0.41$            & 0.03 & $-0.006$            & 0.01 & $\phantom{-}0.34$   & 0.03\\
11	& $\Omegab=0.12$ 		& 51	& 18. & 4.9  &  $\phantom{-}0.37$  & 0.04 & $-0.02$             & 0.01 & $\phantom{-}0.40$   & 0.02\\
12	& $\Zmet=0.0\Zsol$ 		& 9	& 19. & 1.7  &  $-0.56$            & 0.04 & $\phantom{-}0.004$  & 0.01 & $\phantom{-}0.10$   & 0.04\\
12	& $\Zmet=1.0\Zsol$ 		& 51	& 14. & 4.7  &  $\phantom{-}0.05$  & 0.03 & $\phantom{-}0.0$    & 0.02 & $\phantom{-}0.03$   & 0.01\\
13	& $T_{\rm i}=0\kelvin$ 		& 51	& 18. & 6.7  &  $\phantom{-}0.00$  & 0.02 & $\phantom{-}0.0$    & 0.01 & $\phantom{-}0.003$  & 0.01\\
13	& $T_{\rm i}=10^7\kelvin$	& 49	& 41. & 18.  &  $-0.02$            & 0.02 & $-0.01$             & 0.03 & $\phantom{-}0.002$  & 0.03\\

\end{tabular}
\label{tab:scatterstats}
\end{center}
\end{table*}

%% file: paper.bbl
\begin{thebibliography}{}
\bibitem[\protect\citename{Anninos \& Norman }1996]{AN96} Anninos P., Norman M.L., 1996, \ApJ, 459, 12
\bibitem[\protect\citename{Benz }1990]{BENZ} Benz W., 1990, in Buchler J. R., ed., Numerical Modelling of Stellar Pulsations: Problems and Prospects. Kluwer, Dordrecht, p. 269
\bibitem[\protect\citename{Binney \& Tremaine }1987]{BT} Binney J.,Tremaine S., 1987, Galactic Dynamics, Princeton Series in Astrophysics
\bibitem[\protect\citename{Bond \& Efstathiou }1984]{BE} Bond J.R., Efstathiou G., 1984, \ApJ, 285, L45
\bibitem[\protect\citename{Bryan \& Norman }1998]{BN98} Bryan G.R., Norman M.L., 1998, \ApJ, 495, 80
\bibitem[\protect\citename{Cen \etal }1994]{CEN94} Cen R., Miralda-Escud\'e J., Ostriker J.P., Rauch M., 1994, \ApJ, 437, L9
\bibitem[\protect\citename{Cen \etal }1998]{CEN98} Cen R., Phelps S., Miralda--Escud\'e J., Ostriker J.P., 1998, \ApJ, 496, 577
\bibitem[\protect\citename{Cen \& Ostriker }1994]{CO94} Cen R., Ostriker J., 1994, \ApJ, 429, 4
\bibitem[\protect\citename{Cen \& Ostriker }1992]{CO92} Cen R., Ostriker J., 1992, \ApJ, 393, 22
\bibitem[\protect\citename{Centrella \& Melott }1983]{CM83} Centrella J., Melott A.L., 1983, \Nature, 305, 196
\bibitem[\protect\citename{Cole }1991]{Cole91} Cole S., 1991, \ApJ, 367, 45
\bibitem[\protect\citename{Copi, Schramm \& Turner }1995]{CST} Copi C.J., Schramm D.N., Turner M.S., 1995, \ApJ, 455, 95
\bibitem[\protect\citename{Couchman }1991]{COUCH} Couchman H.M.P., 1991, \ApJ, 368, L23
\bibitem[\protect\citename{Couchman, Thomas \& Pearce }1995]{CTP} Couchman H.M.P., Thomas P.A., Pearce F.R., 1995, \ApJ, 452, 797 (CTP95)
\bibitem[\protect\citename{Davis \etal }1985]{DEFW} Davis M., Efstathiou G., Frenk C.S., White S.D.M., 1985, \ApJ, 292, 371
\bibitem[\protect\citename{Efstathiou \etal }1985]{EDFW} Efstathiou G., Davis M., Frenk C.S., White S.D.M., 1985, \ApJS, 57, 241
\bibitem[\protect\citename{Efstathiou \etal }1988]{EDFW88} Efstathiou G., Davis M., Frenk C.S., White S.D.M., 1988, \MNRAS, 235, 715
\bibitem[\protect\citename{Eke, Cole \& Frenk }1996]{ECF} Eke V.R., Cole S.M., Frenk C.S., 1996, \MNRAS, 282, 263
\bibitem[\protect\citename{Eke, Navarro \& Frenk }1998]{ENF98} Eke V.R., Navarro J.F., Frenk C.S., 1998, \ApJ, 503, 569
\bibitem[\protect\citename{Evrard }1990]{Evrard90} Evrard A.E., 1990, \ApJ, 363, 349
\bibitem[\protect\citename{Evrard, Summers \& Davis }1994]{ESD} Evrard A.E., Summers F.J., Davis M., 1994, \ApJ, 422, 11
\bibitem[\protect\citename{Frenk \etal }1999]{FRENK99} Frenk C.S. \etal, 1999, \ApJ, in press
\bibitem[\protect\citename{Gingold \& Monaghan }1977]{GM77} Gingold R.A., Monaghan J.J., 1977, \MNRAS, 181, 375
\bibitem[\protect\citename{Hernquist \etal }1996]{HKWE96} Hernquist L., Katz N., Weinberg D.H., Miralda-Escud\'e J., 1996, \ApJ, 457, L57
\bibitem[\protect\citename{Hockney \& Eastwood }1981]{HOCKEAST} Hockney R.W., Eastwood J.W., 1981, Computer Simulation Using Particles, McGraw-Hill
\bibitem[\protect\citename{Katz, Hernquist \& Weinberg }1999]{KHW99} Katz N., Hernquist L., Weinberg D.H., 1999, Highly Redshifted Radio Lines, ASP Conf. Series Vol. 156, Ed. by C. L. Carilli, S. J. E. Radford, K. M. Menten, \& G. I. Langston, p.1
\bibitem[\protect\citename{Katz, Hernquist \& Weinberg }1992]{KHW} Katz N., Hernquist L., Weinberg D.H., 1992, \ApJ, 399, L109
\bibitem[\protect\citename{Katz }1992]{K92} Katz N., 1992, \ApJ, 391, 502
\bibitem[\protect\citename{Klypin \& Shandarin }1983]{KlypShand} Klypin A.A., Shandarin S.F., 1983, \MNRAS, 204, 891
\bibitem[\protect\citename{Lacey \& Cole }1994]{LC94} Lacey C., Cole S., 1994, \MNRAS, 271, 676
\bibitem[\protect\citename{Lucy }1977]{L77} Lucy L., 1977, \AJ, 82, 1013
\bibitem[\protect\citename{Monaghan \& Gingold }1983]{MONGING} Monaghan J.J., Gingold R.A., 1983, J. Comput. Phys., 52, 374
\bibitem[\protect\citename{Monaghan \& Lattanzio }1986]{MONLAT} Monaghan J.J., Lattanzio J.C., 1986, \AA, 158, 207
\bibitem[\protect\citename{Monaghan }1992]{MON} Monaghan J.J., 1992, \ARAA, 30, 543
\bibitem[\protect\citename{Mushotzky \etal }1996]{MLATFMKH} Mushotzky R., Loewenstein M., Arnaud K.A., Tamura T., Fukazawa Y., Matsushita K., Kikuchi K., Hatsukade I., 1996, \ApJ, 466, 686
\bibitem[\protect\citename{Navarro \& Steinmetz }1997]{NS97} Navarro J.F., Steinmetz M., 1997, \ApJ, 478, 13
\bibitem[\protect\citename{Navarro, Frenk \& White }1995]{NFW95} Navarro J.F., Frenk C.S., White S.D.M., 1995, \MNRAS, 275, 720
\bibitem[\protect\citename{Navarro \& White }1993]{NW93} Navarro J.F., White S.D.M., 1993, \MNRAS, 265, 271
\bibitem[\protect\citename{Pearce \etal }1999]{VGF1} Pearce F.R., Jenkins A., Frenk C.S., Thomas P.A., Colberg J.M., White S.D.M., Couchman H.M.P., Peacock J.A., Efstathiou G., Nelson A.H., 1999, accepted ApJL
\bibitem[\protect\citename{Pearce \& Couchman }1997]{PC} Pearce F.R., Couchman H.M.P., 1997, New Astronomy, 2, 411
\bibitem[\protect\citename{Raymond, Cox \& Smith }1976]{RCS} Raymond J.C., Cox D.P., Smith B.W., 1976, \ApJ, 204, 290
\bibitem[\protect\citename{Steinmetz \& White }1997]{SW} Steinmetz M., White S.D.M., 1997, \MNRAS, 288, 545
\bibitem[\protect\citename{Steinmetz \& M\"uller }1995]{SM95} Steinmetz M., M\"uller E., 1995, \MNRAS, 276, 549
\bibitem[\protect\citename{Sutherland \& Dopita }1993]{SUTDOP} Sutherland R.S., Dopita M.A., 1993, \ApJS, 88, 253
\bibitem[\protect\citename{Thacker \etal }1998]{THACKER} Thacker R.J., Tittley E.R., Pearce F.R., Couchman H.M.P., Thomas P.A., 1998, submitted \MNRAS (astro-ph/9809221)
\bibitem[\protect\citename{Theuns \etal }1998]{THEUNS98} Theuns T., Leonard A., Efstathiou G., Pearce F.R., Thomas P.A., 1998, \MNRAS, 301, 478
\bibitem[\protect\citename{Thomas \& Couchman }1992]{THOMCOUCH} Thomas P.A., Couchman H.M.P., 1992, \MNRAS, 257, 11 (TC92)
\bibitem[\protect\citename{Vianna \& Liddle }1996]{VL} Viana P.T.P., Liddle A.R., 1996, \MNRAS, 281, 323
\bibitem[\protect\citename{Weinberg, Hernquist \& Katz }1997]{WHK97} Weinberg D.H., Hernquist L., Katz N., 1997, \ApJ, 477, 8
\bibitem[\protect\citename{White, Efstathiou \& Frenk }1993]{WEF} White S.D.M., Efstathiou G., Frenk C.S., 1993, \MNRAS, 262, 1023
\bibitem[\protect\citename{White \& Frenk }1991]{WF} White S.D.M., Frenk C.S., 1991, \ApJ, 379, 52
\bibitem[\protect\citename{White \& Rees }1978]{WR} White S.D.M., Rees M.J., 1978, \MNRAS, 183, 341
\bibitem[\protect\citename{Wood }1981]{Wood} Wood D., 1981, \MNRAS, 194, 201
\end{thebibliography}
